\documentclass[twocolumn,showpacs,superscriptaddress,amsmath,amssymb,prc,preprintnumbers]{revtex4-1}
\usepackage[pdftex]{graphicx}
\usepackage{dcolumn}
\usepackage{bm}
\usepackage{ifpdf}
\usepackage{epstopdf}
\usepackage{slashed}
\usepackage{amsfonts}
\usepackage{mathrsfs}
\usepackage{amssymb}
\usepackage{multirow}
\usepackage{CJK}
\usepackage[pdftex]{xcolor}
\usepackage{changes}
\def\brakket#1#2#3{\left\langle{#1}\middle|{#2}\middle|{#3}\right\rangle}
\def\nuc#1#2#3{{}^{#2}_{#3}\mathrm{#1}}
\begin{document}
\begin{CJK*}{UTF8}{}
  %
  \title{Tensor-force effects on shell-structure evolution in $N = 82$ isotones and $Z = 50$ isotopes in the relativistic Hartree-Fock theory}
  \author{Zhiheng Wang (\CJKfamily{gbsn}{王之恒})}
  \affiliation{School of Nuclear Science and Technology, Lanzhou University, Lanzhou 730000, China}
  \affiliation{Joint Department for Nuclear Physics, Lanzhou University and Institute of Modern Physics,
    Chinese Academy of Sciences, Lanzhou 730000, China}
  \author{Tomoya Naito (\CJKfamily{gbsn}{内藤智也})}
  \affiliation{Department of Physics, Graduate School of Science, The University of Tokyo,
    Tokyo 113-0033, Japan}
  \affiliation{RIKEN Nishina Center, Wako 351-0198, Japan}

  \author{Haozhao Liang (\CJKfamily{gbsn}{梁豪兆})}
  \affiliation{Department of Physics, Graduate School of Science, The University of Tokyo,
    Tokyo 113-0033, Japan}
  \affiliation{RIKEN Nishina Center, Wako 351-0198, Japan}

  \date{\today}
  \begin{abstract}
    The evolutions of the energy difference between the neutron states $1i_{13/2}$ and $1h_{9/2}$ in the $N=82$ isotones and
    that between the proton states $1h_{11/2}$ and $1g_{7/2}$ in the $Z = 50$ isotopes are investigated within the framework of the relativistic Hartree-Fock theory, using the density-dependent effective interactions PKA1, PKO$i$ ($i = 1 $, $ 2 $, $ 3 $), and a new interaction developed in this study.
    By identifying the contributions of the tensor force, which is naturally induced via the Fock terms, we find that the tensor force plays crucial roles in the evolution of the shell structure.
    The strength of the tensor force is also explored. It is found that moderately increasing the coupling strength of pion-nucleon coupling, i.e., $f_{\pi}$, will significantly improve the description of the shell-structure evolution.
    In particular, reducing the density dependence of $f_{\pi}$ is shown to be preferable, in comparison to enlarging $f_{\pi}$ with a factor.
    This is consistent with the idea of ``tensor renormalization persistency'' and provides valuable guidance for the development of the nuclear energy density functional in the relativistic framework.
  \end{abstract}
  \maketitle
\end{CJK*}
\section{Introduction}
\par
Shell structure belongs to the most important properties of nuclear systems.
By introducing strong spin-orbit coupling to the harmonic oscillator potential,
Haxel, Jensen, and Suess~\cite{
  Haxel1949Phys.Rev.75.1766}
and Mayer~\cite{
  Mayer1949Phys.Rev.75.1969}
successfully explained the magic numbers associated with the shell closure.
Owing to the development of modern radioactive
nuclear beam facilities and experimental detectors~\cite{
  Tanihata1985Phys.Lett.B160.380,
  Tanihata1985Phys.Rev.Lett.55.2676,
  Gade2008Prog.Part.Nucl.Phys.60.161,
  Pfuetzner2012Rev.Mod.Phys.84.567,
  Tanihata2013Prog.Part.Nucl.Phys.68.215,
  Campbell2016Prog.Part.Nucl.Phys.86.127,
  Wakasa2017Prog.Part.Nucl.Phys.96.32,
  Nakamura2017Prog.Part.Nucl.Phys.97.53},
the landscape of nuclei extends from the $\beta$-stability line to the regime of exotic ones with unbalanced $N/Z$ ratio~\cite{
  Thoennessen2013Rep.Prog.Phys.76.056301,
  Wang2017Chin.Phys.C41.030003}.
Meanwhile, traditional shell structure does not remain solid.
The modification of shell closures has become one of the most intriguing issues in recent decades~\cite{
  Sorlin2008Prog.Part.Nucl.Phys.61.602,
  Otsuka2020Rev.Mod.Phys.92.015002}.
\par
The shell evolution, which leads to the appearance of new magic numbers as well as the disappearance of traditional ones, challenges the traditional understanding of nuclear physics~\cite{
  Otsuka2001Phys.Rev.Lett.87.082502,
  Wienholtz2013Nature498.346,
  Steppenbeck2013Nature502.207}.
For example, the tensor force~\cite{
  Rarita1941Phys.Rev.59.556,
  Gerjuoy1942Phys.Rev.61.138},
which had been neglected for a long time in the effective interactions~\cite{
  Bender2003Rev.Mod.Phys.75.121},
has regained tremendous interest, mainly because of its characteristic effects on the spin-orbit splitting and hence on the shell evolution~\cite{
  Otsuka2005Phys.Rev.Lett.95.232502}.
It has been pointed out that the tensor force is repulsive between the two nucleons which are both spin up ($j=l+1/2$) or spin down ($j=l-1/2$),
while it is attractive when one nucleon is spin-up and the other is spin-down~\cite{
  Otsuka2005Phys.Rev.Lett.95.232502}.
Such a character of spin dependence affects the spin-orbit splitting, especially for the nuclides located in the exotic regime.
\par
During the last two decades, the tensor-force effects on the single-particle energies have been studied extensively, in both the nuclear density functional theory (DFT)~\cite{
  Nakada2003Phys.Rev.C68.014316,
  Otsuka2006Phys.Rev.Lett.97.162501,
  Brink2007Phys.Rev.C75.064311,
  Lesinski2007Phys.Rev.C76.014312,
  Long2008Europhys.Lett.82.12001,
  Zou2008Phys.Rev.C77.014314,
  Nakada2008Phys.Rev.C78.054301,
  Moreno-Torres2010Phys.Rev.C81.064327,
  Wang2011Phys.Rev.C83.054305,
  Wang2011Phys.Rev.C84.044333,
  Dong2011Phys.Rev.C84.014303,
  Anguiano2012Phys.Rev.C86.054302,
  Sagawa2014Prog.Part.Nucl.Phys.76.76,
  Liang2015Phys.Rep.570.1,
  Shi2017Phys.Rev.C95.034307,
  Shen2019Phys.Rev.C99.034322,
  Dong2020Phys.Rev.C101.014305,
  Nakada2020Int.J.Mod.Phys.E29.1930008}
and the shell model~\cite{
  Otsuka2001Phys.Rev.Lett.87.082502,
  Otsuka2005Phys.Rev.Lett.95.232502,
  Brown2006Phys.Rev.C74.061303,
  Stanoiu2008Phys.Rev.C78.034315,
  Kaneko2011Phys.Rev.C83.014320,
  Tsunoda2014Phys.Rev.C89.031301,
  Chen2019Phys.Rev.Lett.122.212502}.
\par
In spite of this, the strength of the in-medium effective tensor force and even its sign are still under discussion~\cite{
  Sagawa2014Prog.Part.Nucl.Phys.76.76,
  Otsuka2020Rev.Mod.Phys.92.015002}.
Traditionally, the parameters of the effective interactions are fitted to the bulk properties, which are shown to be not sensitive enough to the tensor force.
As a result, the properties of the in-medium effective tensor force are far away from being efficiently constrained.
Actually, pinning down the nature of the tensor force is one of the crucial aspects for the ultimate understanding of the effective nuclear interactions employed in, for example, the DFT~\cite{
  Otsuka2006Phys.Rev.Lett.97.162501,
  Lalazissis2009PRC80.041301(R),
  Cao2009Phys.Rev.C80.064304,
  Bai2011Phys.Rev.C83.054316,
  Anguiano2011Phys.Rev.C83.064306,
  Afanasjev2015Phys.Rev.C92.044317,
  Shen2019Phys.Rev.C99.034322}.
To achieve this, one needs to find the observables that are sensitive to the tensor force but not so much to the other components of nuclear force.
The single-particle energies, which are largely affected by the spin-orbit splitting, can serve as such kind of benchmarks.
It is thus of great significance to study how the tensor force affects the single-particle energies, especially the shell-structure evolution along the isotopic or isotonic chains.
\par
In 2004, by neutron and proton transfer reactions, Schiffer \textit{et al.}~\cite{
  Schiffer2004Phys.Rev.Lett.92.162501}
investigated the differences between the energies of the neutron ($\nu$) states $1i_{13/2}$ and $1h_{9/2}$ in $N=82$ isotones and those of the proton ($\pi$) states $1h_{11/2}$ and $1g_{7/2}$ in $Z = 50$ isotopes.
This experimental progress immediately attracted a lot of interest in theoretical research and became a popular playground for investigating the shell evolution and the hidden mechanisms.
In particular, Col\`{o} \textit{et al.}~\cite{
  Colo2007Phys.Lett.B646.227}
found that the introduction of the tensor force on top of the Skyrme effective interactions can fairly well reproduce the isospin dependence of the energy differences in the above-mentioned isotones and isotopes, while the effective interactions without the explicit tensor force fail.
The tensor-force effects were also explored within the relativistic framework, i.e., the relativistic Hartree-Fock (RHF) theory~\cite{
  Long2008Europhys.Lett.82.12001}.
It was found that the tensor force arising from the pion exchange plays a crucial role in explaining the evolution of the energy difference with the neutron or proton number.
\par
The tensor force can be added on top of the existing Skyrme interactions perturbatively and refitted individually; the $ \text{SLy5} + \text{T} $~\cite{
  Colo2007Phys.Lett.B646.227}
and $ \text{SIII} + \text{T} $~\cite{
  Brink2007Phys.Rev.C75.064311}
functionals were established in this way.
An alternative way, which is practically more accurate and more systematic, is to fit the tensor force on the same footing with the other terms, and the representative functionals are
the Skxta and Skxtb~\cite{
  Brown2006Phys.Rev.C74.061303},
the T$IJ$ family~\cite{
  Lesinski2007Phys.Rev.C76.014312},
$ \text{SkP}_{\text{T}} $, $ \text{SLy4}_{\text{T}} $, and $ \text{SkO}_{\text{T}} $~\cite{
  Zalewski2008Phys.Rev.C77.024316},
and SAMi-T~\cite{
  Shen2019Phys.Rev.C99.034322}.
Similar strategies to introduce the tensor force have also been applied for the finite-range Gogny and M3Y interactions~\cite{
  Otsuka2006Phys.Rev.Lett.97.162501,
  Nakada2020Int.J.Mod.Phys.E29.1930008,
  Brown2006Phys.Rev.C74.061303}.
Nevertheless, both of these two ways inevitably increase the number of free parameters due to the additional inclusion of the tensor force, which may be one of the reasons why the tensor force is neglected in most of the effective interactions in the nonrelativistic framework.
\par
In the covariant density functional theory (CDFT) with exchange terms, namely the RHF theory, the strong spin-orbit coupling is treated naturally and the tensor force is consistently contained without extra free parameters via the Fock terms of the relevant meson-nucleon interactions
~\cite{
  Long2006Phys.Lett.B640.150,
  Long2007Phys.Rev.C76.034314,
  Liang2008Phys.Rev.Lett.101.122502,
  Liang2012Phys.Rev.C85.064302,
  Niu2017Phys.Rev.C95.044301}.
On the other hand, the tensor force is mixed together with the other components, e.g., the spin-orbit force and other central forces.
This leads to great difficulties in the quantitative analysis of the tensor force in CDFT, despite the noticeable progress on the tensor-force effects arising from the $\pi$-pseudovector and $\rho$-tensor couplings~\cite{
  Long2008Europhys.Lett.82.12001,
  Wang2013Phys.Rev.C87.047301,
  Li2014Phys.Lett.B732.169,
  Li2016Phys.Lett.B753.97,
  Li2016Phys.Rev.C93.054312,
  Li2019Phys.Lett.B788.192}.
In particular, a set of formulas with Lorentz covariance were developed to describe the spin-dependent nature of the nuclear force within the RHF theory in 2015, and they could reproduce the spin dependence of the two-body interaction matrix elements quite well~\cite{
  Jiang2015Phys.Rev.C91.025802,
  Jiang2015Phys.Rev.C91.034326,
  Zong2018Chin.Phys.C42.024101}.
However, these formulas are not straightforwardly related to the tensor force in the conventional sense.
Thus, the corresponding results cannot be compared directly with the tensor-force effects calculated by the nonrelativistic DFT.
Aiming at a direct comparison between the tensor-force effects in the CDFT and those in the nonrelativistic DFT, recently, the tensor force in each meson-nucleon coupling has been identified within the RHF theory~\cite{
  Wang2018Phys.Rev.C98.034313}.
In addition, a method to quantitatively evaluate the contributions of the tensor forces arising from these meson-nucleon couplings has also been illustrated.
\par
In this work, we will investigate the isospin evolution of the single-particle energy differences in the $N = 82$ isotones and the $Z = 50$ isotopes mentioned above within the RHF theory.
In particular, the effects of the tensor force will be studied with the method newly developed in Ref.~\cite{
  Wang2018Phys.Rev.C98.034313}.
The strength of the tensor force will be explored, aiming to provide guidance for the future development of the effective interactions.
To this end, it is noticed that the single-particle energies, as well as the collective excitations, are affected by various ``dynamic correlations'', among which the particle-vibration coupling (PVC)~\cite{
  Bertsch1983Rev.Mod.Phys.55.287,
  Litvinova2006Phys.Rev.C73.044328,
  Litvinova2011Phys.Rev.C84.014305,
  Niu2012Phys.Rev.C85.034314,
  Cao2014Phys.Rev.C89.044314,
  Niu2014Phys.Rev.C90.054328,
  Niu2015Phys.Rev.Lett.114.142501,
  Karakatsanis2017Phys.Rev.C95.034318}
is considered to play the most important role, especially for the spherical nuclei.
Thus, the experimental data of the single-particle energy are not expected to be directly
comparable with the corresponding results calculated by DFT, which are purely mean-field-level quantities.
\par
In Sec.~\ref{sec:theory}, we briefly introduce the framework of RHF theory, as well as the main ideas to extract the contributions of the tensor force in the RHF theory.
The results will be discussed in Sec.~\ref{Sec:results}.
A summary and perspectives will be given in Sec.~\ref{sec:summary}.
\section{Theoretical Framework}
\label{sec:theory}
\par
In this section, the basic framework of the RHF theory will be recalled, and more details can be found in Refs.~\cite{
  Bouyssy1987Phys.Rev.C36.380,
  Shi1995Phys.Rev.C52.144,
  Long2006Phys.Lett.B640.150,
  Long2007Phys.Rev.C76.034314,
  Sun2008Phys.Rev.C78.065805,
  Long2010Phys.Rev.C81.024308,
  Ebran2011Phys.Rev.C83.064323,
  Li2018Eur.Phys.J.A54.133,
  Geng2019Phys.Rev.C100.051301,
  Geng2020Phys.Rev.C101.064302}.
The method to extract the contributions of the tensor force will also be briefly presented.
\subsection{Relativistic Hartree-Fock theory}
\par
In the relativistic framework, the nucleons interact with each other via the exchange of mesons and photons~\cite{
  Walecka1974Ann.Phys.83.491,
  Ring1996Prog.Part.Nucl.Phys.37.193,
  Bender2003Rev.Mod.Phys.75.121,
  Vretenar2005Phys.Rep.409.101,
  Meng2006Prog.Part.Nucl.Phys.57.470,
  Niksic2011Prog.Part.Nucl.Phys.66.519,
  Meng2015J.Phys.G:Nucl.Part.Phys.42.093101}.
Based on this picture, the starting point of the RHF theory is a standard Lagrangian density, which contains the degrees of freedom associated with the nucleon field, various meson fields, and the photon field.
Through the Legendre transformation, the Hamiltonian of the system can be derived. With the equations of motion for the mesons and photon, the Hamiltonian can be expressed only with the degree of freedom of the nucleon field, and reads
\begin{align}
  H
  = & \,
      \int
      d^3 x \,
      \bar{\psi} \left( x \right)
      \left[
      - i \bm{\gamma} \cdot \bm{\nabla} + M
      \right]
      \psi \left( x \right)
      \notag \\
    & \,
      +
      \frac{1}{2}
      \sum_{\phi}
      \iint
      d^3 x \, d^4 y \,
      \bar{\psi} \left( x \right)
      \bar{\psi} \left( y \right)
      \Gamma_{\phi} \left( x, y \right)
      D_{\phi} \left( x, y \right)
      \notag \\
    & \,
      \qquad
      \times
      \psi \left( y \right)
      \psi \left( x \right),
      \label{Hamil}
\end{align}
where $\phi$ denotes the meson-nucleon couplings,
including here the Lorentz $\sigma$-scalar ($\sigma$-S),
$\omega$-vector ($\omega$-V),
$\rho$-vector ($\rho$-V),
$\rho$-tensor ($\rho$-T),
$\rho$-vector-tensor ($\rho$-VT),
and $\pi$-pseudovector ($\pi$-PV) couplings,
as well as the photon-vector ($A$-V) coupling.
\par
The interaction vertices $ \Gamma_{\phi} \left( x, y \right) $ in the Hamiltonian~\eqref{Hamil} read
\begin{subequations}
  \label{eq:vertex}
  \begin{align}
    \Gamma_{\text{$ \sigma $-S}} \left( x, y \right)
    = & \,
        -
        \left[ g_{\sigma} \right]_x
        \left[ g_{\sigma} \right]_y, \\
    \Gamma_{\text{$ \omega $-V}} \left( x, y \right)
    = & \,
        +
        \left[ g_{\omega} \gamma_{\mu} \right]_x
        \left[ g_{\omega} \gamma^{\mu} \right]_y, \\
    \Gamma_{\text{$ \rho $-V}} \left( x, y \right)
    = & \,
        +
        \left[ g_{\rho} \gamma_{\mu} \vec{\tau} \right]_x
        \cdot
        \left[ g_{\rho} \gamma^{\mu} \vec{\tau} \right]_y, \\
    \Gamma_{\text{$ \rho $-T}} \left( x, y \right)
    = & \,
        +
        \left[ \frac{f_{\rho}}{2M} \sigma_{\mu \nu} \vec{\tau} \partial^{\nu} \right]_x
        \cdot
        \left[ \frac{f_{\rho}}{2M} \sigma^{\mu \lambda} \vec{\tau} \partial_{\lambda} \right]_y,
        \label{rhot} \\
    \Gamma_{\text{$ \rho $-VT}} \left( x, y \right)
    = & \,
        +
        \left[ \frac{f_{\rho}}{2M} \sigma_{\mu \nu} \vec{\tau} \partial^{\mu} \right]_x
        \cdot
        \left[ g_{\rho} \gamma^{\nu} \vec{\tau} \right]_y
        +
        \text{($ x \leftrightarrow y $)},
        \label{rhvt} \\
    \Gamma_{\text{$ \pi $-PV}} \left( x, y \right)
    = & \,
        -
        \left[ \frac{f_{\pi}}{m_{\pi}} \vec{\tau} \gamma_5 \gamma_{\mu} \partial^{\mu} \right]_x
        \cdot
        \left[ \frac{f_{\pi}}{m_{\pi}} \vec{\tau} \gamma_5 \gamma_{\nu} \partial^{\nu} \right]_y,
        \label{pi-sv} \\
    \Gamma_{\text{$ A $-V}} \left( x, y \right)
    = & \,
        +
        \left[ e \gamma_{\mu} \frac{1 - \tau_3}{2} \right]_x
        \left[ e \gamma^{\mu} \frac{1 - \tau_3}{2} \right]_y,
  \end{align}
\end{subequations}
with the meson-nucleon coupling strengths $g_{\phi}$ and $f_{\phi}$, the nucleon mass $M$, and the meson masses $m_{\phi}$.
If the retardation effect is neglected~\cite{
  Bouyssy1987Phys.Rev.C36.380},
the meson and photon propagators, $ D_{\phi} \left( x, y \right) $,
become the standard Yukawa and Coulomb forms,
\begin{subequations}
  \begin{align}
    D_{\phi} \left( x, y \right)
    & =
      \frac{1}{4 \pi}
      \frac{e^{-m_{\phi} \left| \bm{r}_1 - \bm{r}_2 \right|}}{\left| \bm{r}_1 - \bm{r}_2 \right|}, \\
    D_{\text{$ A $-V}} \left( x, y \right)
    & =
      \frac{1}{4 \pi}
      \frac{1}{\left| \bm{r}_1 - \bm{r}_2 \right|},
  \end{align}
\end{subequations}
respectively.
Hereafter, we use $\bm{r}_1$ and $\bm{r}_2$ to denote the spatial coordinates at vertices $x$ and $y$, and the indices ``1'' and ``2'' are always used to denote the vertices.
\par
The meson-nucleon coupling strengths are taken as functions of the baryonic density.
For the convenience of the following discussions about the strength of the tensor force, here we explicitly present the density dependence of the $\pi$-PV coupling $f_{\pi}$, which reads
\begin{equation}
  \label{Eq:DDPI}
  f_{\pi} \left( \rho_{\text{b}} \right)
  =
  f_{\pi} \left( 0 \right)
  e^{-a_{\pi} \xi},
\end{equation}
where $ \xi = \rho_{\text{b}} / \rho_{\text{sat.}} $, $ \rho_{\text{sat.}}$ denotes the saturation density of the nuclear matter, and $ f_{\pi} \left( 0 \right) $ corresponds to the coupling strength at zero density.
The density dependence of the other meson-nucleon couplings can be found in Refs.~\cite{
  Long2006Phys.Lett.B640.150,
  Long2007Phys.Rev.C76.034314}.
\par
The nucleon-field operators $\psi \left( x \right)$ and $\psi^{\dagger} \left( x \right)$ can be expanded on the set of creation and annihilation operators defined by a complete set of Dirac spinors
$ \left\{ \varphi_{\alpha} \left( \bm{r} \right) \right\} $.
Then, the energy functional can be obtained through the expectation value of the Hamiltonian on the trial Hartree-Fock state {$ \left| \Phi_0 \right\rangle $} under the no-sea approximation~\cite{
  Walecka1974Ann.Phys.83.491}.
It can be expressed as
\begin{widetext}
  \begin{align}
    E
    = & \,
        \brakket{\Phi_0}{H}{\Phi_0}
        -
        AM
        \notag \\
    = & \,
        E^{\text{K}}
        +
        \sum_{\phi}
        \left( E^{\text{D}}_{\phi} + E^{\text{E}}_{\phi} \right)
        \notag \\
    = & \,
        \sum_{\alpha}
        \int
        d \bm{r} \,
        \bar{\varphi}_{\alpha} \left( \bm{r} \right)
        \left( - i \bm{\gamma} \cdot \bm{\nabla} + M \right)
        \varphi_{\alpha} \left( \bm{r} \right)
        -
        AM
        \notag \\
      & \,
        +
        \frac{1}{2}
        \sum_{\phi, \alpha \beta}
        \left\{
        \iint
        d \bm{r}_1 \, d \bm{r}_2 \,
        \bar{\varphi}_{\alpha} \left( \bm{r}_1 \right)
        \bar{\varphi}_{\beta} \left( \bm{r}_2 \right)
        \Gamma_{\phi} \left( \bm{r}_1, \bm{r}_2 \right)
        D_{\phi} \left( \bm{r}_1, \bm{r}_2 \right)
        \varphi_{\alpha} \left( \bm{r}_1 \right)
        \varphi_{\beta} \left( \bm{r}_2 \right)
        \right.
        \notag \\
      & \,
        \qquad
        \left.
        -
        \iint
        d \bm{r}_1 \, d \bm{r}_2 \,
        \bar{\varphi}_{\alpha} \left( \bm{r}_1 \right)
        \bar{\varphi}_{\beta} \left( \bm{r}_2 \right)
        \Gamma_{\phi} \left( \bm{r}_1, \bm{r}_2 \right)
        D_{\phi} \left( \bm{r}_1, \bm{r}_2 \right)
        \varphi_{\beta} \left( \bm{r}_1 \right)
        \varphi_{\alpha} \left( \bm{r}_2 \right)
        \right\},
        \label{HFEDF}
  \end{align}
\end{widetext}
where $E^{\text{K}}$ denotes the kinetic energy, and $E^{\text{D}}_{\phi}$ and $E^{\text{E}}_{\phi}$ correspond to the energy contributions from the direct (Hartree) and exchange (Fock) terms, respectively.
\par
In spherically symmetric systems, the single-particle states, which are Dirac spinors here,  can be specified by a set of quantum numbers
$ \alpha \equiv \left( a, m_{\alpha} \right) \equiv \left( \tau_a, n_a, l_a, j_a, m_{\alpha} \right) $.
Explicitly, the Dirac spinors have the following expression,
\begin{equation}
  \label{DSSN}
  \varphi_{\alpha} \left( \bm{r} \right)
  =
  \frac{1}{r}
  \begin{pmatrix}
    i G_a \left( r \right) \\
    F_a \left( r \right) \hat{\bm{\sigma}} \cdot \hat{\bm{r}}
  \end{pmatrix}
  \mathscr{Y}_{\alpha}
  \left( \hat{\bm{r}} \right)
  \chi_{\frac{1}{2}} \left( \tau_a \right),
\end{equation}
where $\mathscr{Y}_{\alpha} \left( \hat{\bm{r}} \right)$ are the tensor spherical harmonics defined through the coupling of the spherical harmonics and the spin spinors~\cite{
  Varshalovich1988QuantumTheoryofAngularMomentum_WorldScientific}, and $\chi_{\frac{1}{2}} \left( \tau_a \right)$ is the isospinor.
\par
The variation of the energy functional with respect to Dirac spinors leads to the Hartree-Fock equation, which formally reads
\begin{equation}
  \int
  d \bm{r}' \,
  h \left( \bm{r}, \bm{r}' \right)
  \varphi \left( \bm{r}' \right)
  =
  \varepsilon
  \varphi \left( \bm{r} \right),
\end{equation}
where the Lagrangian multiplier $\varepsilon$ is the single-particle energy including the rest mass of the nucleon.
The single-particle Hamiltonian $h \left(\bm{r}, \bm{r}' \right)$ contains
the kinetic energy $h^{\text{K}}$,
the direct local potential $h^{\text{D}}$,
and the exchange nonlocal potential $h^{\text{E}}$:
\begin{subequations}
  \label{Eq:sph}
  \begin{align}
    h^{\text{K}} \left( \bm{r}, \bm{r}' \right)
    & =
      \left[
      \bm{\alpha} \cdot \bm{p}
      +
      \beta M
      \right]
      \delta \left( \bm{r}, \bm{r}' \right), \\
    h^{\text{D}} \left( \bm{r}, \bm{r}' \right)
    & =
      \left[
      \Sigma_{\text{T}} \left( \bm{r} \right) \gamma_5
      +
      \Sigma_0 \left( \bm{r} \right)
      +
      \beta \Sigma_{\text{S}} \left( \bm{r} \right)
      \right]
      \delta \left( \bm{r}, \bm{r}' \right), \\
    h^{\text{E}} \left( \bm{r}, \bm{r}' \right)
    & =
      \begin{pmatrix}
        Y_G \left( \bm{r}, \bm{r}' \right)
        & Y_F \left( \bm{r}, \bm{r}' \right) \\
        X_G \left( \bm{r}, \bm{r}' \right)
        & X_F \left( \bm{r}, \bm{r}' \right)
      \end{pmatrix}.
  \end{align}
\end{subequations}
The tensor force contributes only to the nonlocal self-energies $X_G$, $X_F$, $Y_G$, and $Y_F$.
See Refs.~\cite{
  Bouyssy1987Phys.Rev.C36.380,
  Long2005PhD.Thesis,
  Long2010Phys.Rev.C81.024308}
for the full expressions of these quantities.
\par
Basically, the Hartree-Fock equation shall be solved iteratively. For open-shell nuclei, the paring correlation is treated using the BCS method.
The zero-range density-dependent interaction~\cite{Dobaczewski1996Phys.Rev.C53.2809},
\begin{equation}
  V \left( r_1, r_2 \right)
  =
  V_0
  \delta \left( r_1 - r_2 \right)
  \left[
    1 - \frac{\rho_{\text{b}} \left( r \right)}{\rho_{\text{sat.}}}
  \right],
\end{equation}
is adopted to calculate the pairing matrix elements.

The strength $V_0$ is uniformly chosen as $ -500 \, \mathrm{MeV}~\mathrm{fm}^3 $.
The active pairing space is limited to the single-particle states below $ +10 \, \mathrm{MeV} $.
It can be shown that the shell structure and its evolution in the nuclei studied here are not sensitive to the value of $V_0$ over a wide range~\cite{
  Long2008Europhys.Lett.82.12001}.
\subsection{Contributions of tensor force}
\par
As already mentioned, the tensor force in each meson-nucleon coupling was identified through nonrelativistic reduction~\cite{
  Wang2018Phys.Rev.C98.034313}.
The method to quantitatively evaluate the contributions of the tensor forces was established as well.
Here, we will not go into the details of nonrelativistic reduction, but merely show the tensor components in the relevant meson-nucleon couplings in a uniform expression as
\begin{equation}
  \label{eq:Vtphi_abcd}
  \hat{\mathcal{V}}^{\text{t}}_{\phi}
  =
  \frac{1}{m_{\phi}^2 + \bm{q}^2}
  \mathcal{F}_{\phi}
  S_{12},
\end{equation}
where $\bm{q}$ is the momentum transfer, and $S_{12}$ is the operator of the tensor force in the momentum space, which reads
\begin{equation}
  \label{S12}
  S_{12}
  \equiv
  \left( \bm{\sigma}_1 \cdot \bm{q} \right)
  \left( \bm{\sigma}_2 \cdot \bm{q} \right)
  -
  \frac{1}{3}
  \left( \bm{\sigma}_1 \cdot \bm{\sigma}_2 \right)
  q^2.
\end{equation}
The coefficient $ \mathcal{F}_{\phi} $ associated with a given meson-nucleon coupling reflects the sign and the rough strength of the tensor force, as displayed in Table \ref{Table:F}.
\begin{table}
  \caption{
    Expressions for $ \mathcal{F}_{\phi} $ in Eq.~\eqref{eq:Vtphi_abcd} for each meson-nucleon coupling. $M^*$ is the Dirac mass of nucleons.}
  \label{Table:F}
  \begin{ruledtabular}
    \begin{tabular}{cc|cc}
      $ \phi $ & $ \mathcal{F}_{\phi} $ & $ \phi $ & $ \mathcal{F}_{\phi} $ \\ \hline
      $ \omega $-V & $ \dfrac{g_{\omega} \left( 1 \right) g_{\omega} \left( 2 \right)}{4M^* \left( 1 \right) M^* \left( 2 \right)} $
                                        & $ \pi $-PV & $ - \dfrac{f_{\pi} \left( 1 \right) f_{\pi} \left( 2 \right)}{m^2_{\pi}} $ \\
      $ \rho $-V & $ \dfrac{g_{\rho} \left( 1 \right) g_{\rho} \left( 2 \right)}{4M^* \left( 1 \right) M^* \left( 2 \right)} $
                                        & $ \rho $-T & $ \dfrac{f_{\rho} \left( 1 \right) f_{\rho} \left( 2 \right)}{4M^2} $ \\
               &
                                        & $ \rho $-VT & $ \dfrac{f_{\rho} \left( 1 \right) g_{\rho} \left( 2 \right)}{4MM^* \left( 2 \right)} + \text{($ 1 \leftrightarrow 2 $)} $ \\
    \end{tabular}
  \end{ruledtabular}
\end{table}
\par
To evaluate quantitatively the contributions of the tensor force
$\hat{\mathcal{V}}^{\text{t}}_{\phi}$ in each meson-nucleon coupling, one first needs to calculate its contributions to the two-body interaction matrix elements.
The explicit formulas are found in Appendix C in Ref.~\cite{
  Wang2018Phys.Rev.C98.034313}.
Further, one can calculate the contributions of the tensor force to the nonlocal self-energies and ultimately its contributions to the single-particle energies.
\section{Results and Discussion}
\label{Sec:results}
\par
\subsection{$N = 82$ isotones}\label{Subsec:N82}
\par
\begin{figure*}[tb]
  \includegraphics[width=0.45\textwidth]{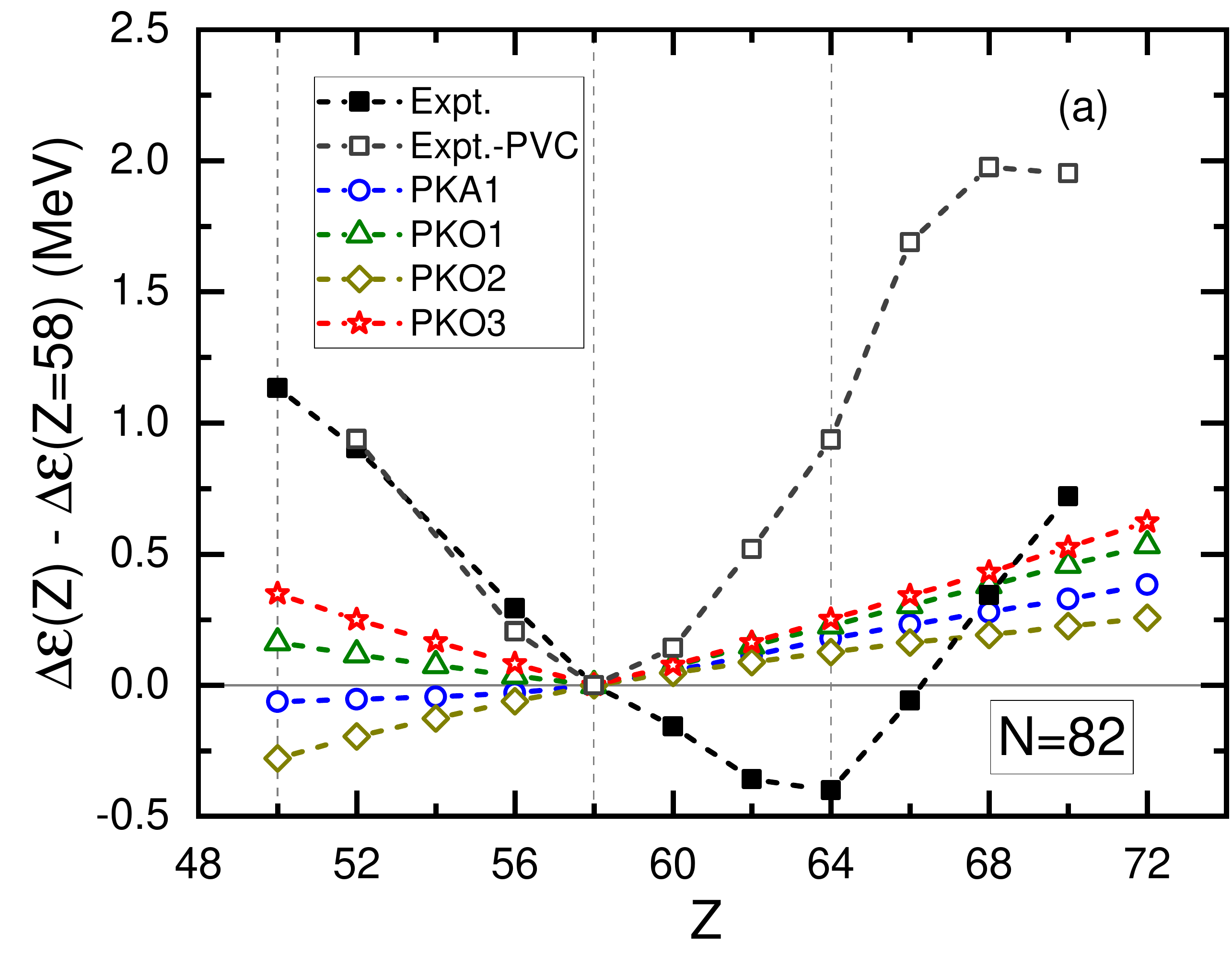}
  \hspace{0.5cm}
  \includegraphics[width=0.45\textwidth]{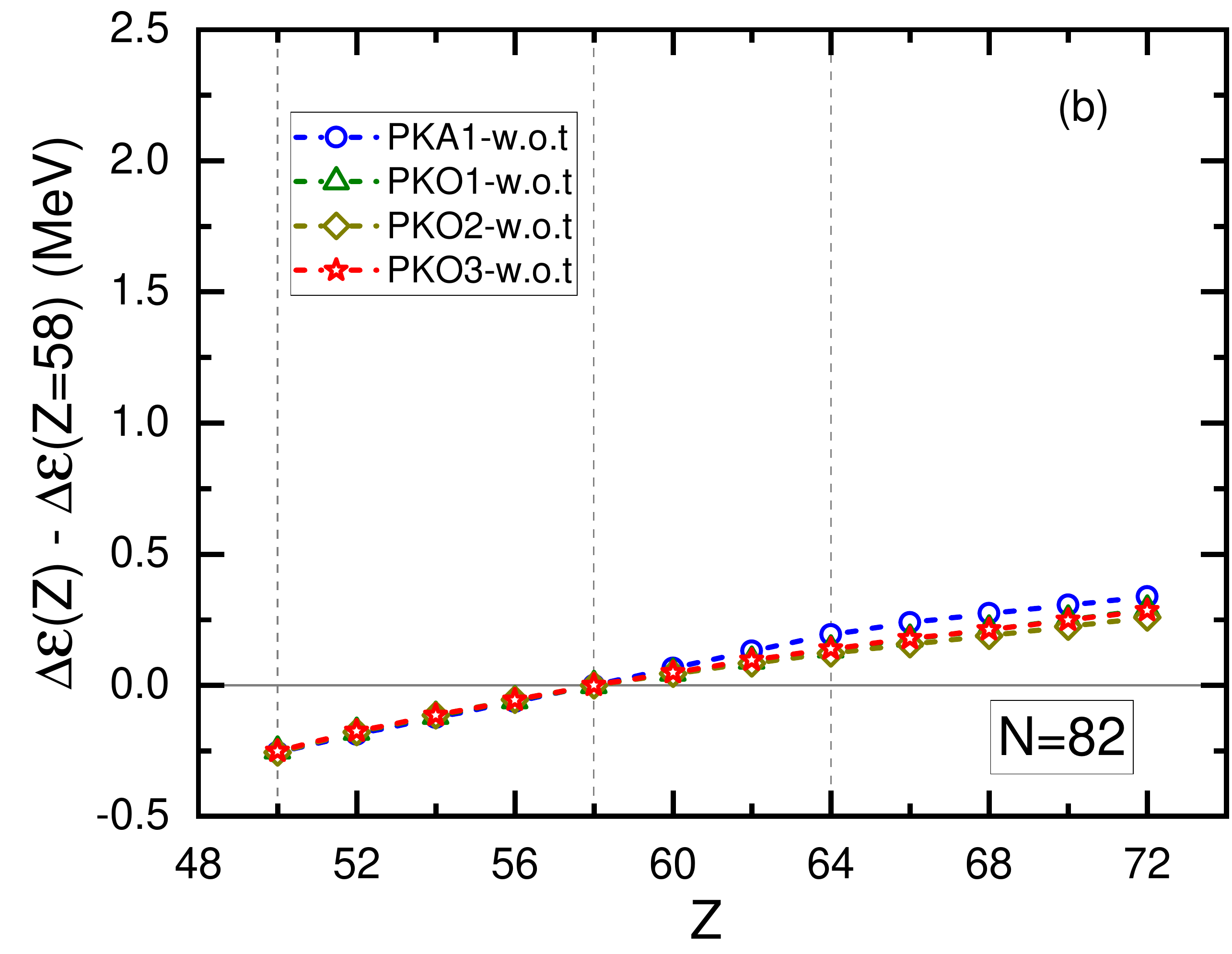}
  \caption{
    Energy differences
    $\Delta \varepsilon \equiv \varepsilon_{\nu 1i_{13/2}}-\varepsilon_{\nu 1h_{9/2}}$
    in the $N=82$ isotones as functions of the proton number.
    Panel~(a) shows the calculated results by the RHF theory with the effective interactions PKA1 \cite{Long2007Phys.Rev.C76.034314} and PKO$i$ ($ i= 1 $, $ 2 $, $ 3 $)~\cite{Long2006Phys.Lett.B640.150, Long2008Europhys.Lett.82.12001}.
    Panel~(b) shows the results of the same calculations but without the tensor-force contributions.
    The pairing is treated with the BCS method.
    The experimental data for comparison include the original ones (filled squares) from Ref.~\cite{
      Schiffer2004Phys.Rev.Lett.92.162501}
    as well as the ones (open squares) in which the correlation of particle-vibration coupling~\cite{
      Afanasjev2015Phys.Rev.C92.044317} is subtracted.
    All the experimental data and the calculated results are normalized with respect to their corresponding values at $Z = 58$.
    See the text for details.}
  \label{Fig:N82-pairing}
\end{figure*}
\par
First, we study the evolution of the energy difference between the states $\nu 1i_{13/2}$ and $\nu 1h_{9/2}$, i.e., $\Delta \varepsilon \equiv \varepsilon_{\nu 1i_{13/2}}-\varepsilon_{\nu 1h_{9/2}}$,
along the $N=82$ isotones.
Shown in Fig.~\ref{Fig:N82-pairing} are the energy differences as functions of the proton number $Z$, normalized with respect to the values at $Z = 58$.
The energies of the single-particle states are determined by various components of the nuclear force, such as the central force and the spin-orbit one.
In contrast, the tensor force mainly affects the evolution of the energy differences between two levels, especially those with opposite spin directions, along the isotopic or isotonic chains \cite{
  Otsuka2020Rev.Mod.Phys.92.015002,
  Otsuka2005Phys.Rev.Lett.95.232502,
  Sagawa2014Prog.Part.Nucl.Phys.76.76,
  Colo2007Phys.Lett.B646.227}.
Because of this, the evolution of the energy difference, rather than the single-particle energies themselves, is usually chosen to benchmark the tensor force.
Since we are interested in the evolution, including mainly the trend and the slope, it is convenient to normalize the theoretical and experimental data with respect to their corresponding values at $ Z= 58$ following the previous choice in Ref.~\cite{Long2008Europhys.Lett.82.12001}.
\par
The original experimental value of $\Delta \varepsilon \left( {Z = 58} \right)$ is $-0.05 \, \mathrm{MeV} $~\cite{
  Schiffer2004Phys.Rev.Lett.92.162501}
and the plot is thus moved upwards by $0.05 \, \mathrm{MeV}$, as shown in Fig.~\ref{Fig:N82-pairing}(a).
It can be seen that the original experimental data present a distinct kink at $Z=64$.
Because the experimental data of single-particle energies contain the beyond-mean-field correlations, they are not supposed to be compared directly with the corresponding calculated results based on the pure mean-field approximation, as emphasized by a series of studies~\cite{
  Colo2007Phys.Lett.B646.227,
  Lesinski2007Phys.Rev.C76.014312,
  Long2008Europhys.Lett.82.12001,
  Afanasjev2015Phys.Rev.C92.044317,
  Shen2018Phys.Lett.B778.344,
  Sagawa2014Prog.Part.Nucl.Phys.76.76}.
Among various beyond-mean-field correlations, the PVC is considered to play the most important role. To fulfill the requirement of the self-consistency, we should take into account the PVC effects based on the RHF theory ($ \text{RHF} + \text{PVC} $).
However, to our knowledge, there is still no accessible $ \text{RHF} + \text{PVC} $ calculation.
On the other hand, we notice that the quasiparticle PVC (QPVC) calculation based on the RMF theory using the nonlinear effective interaction NL3* was performed for the $N=82$ isotones~\cite{
  Afanasjev2015Phys.Rev.C92.044317}.
It was also suggested that the effective tensor force has to be quenched as compared with the earlier estimates without considering the PVC effects~\cite{
  Afanasjev2015Phys.Rev.C92.044317}.
Therefore, here we extract the difference between the results of RMF and QPVC, and deem it, to some extent, the PVC effects.
By excluding the PVC effects from the original data~\cite{
  Schiffer2004Phys.Rev.Lett.92.162501},
we get the corresponding ``pseudodata,''
which are denoted by the open squares in Fig.~\ref{Fig:N82-pairing}(a).
\par
To avoid confusion, we stress that it is far from certain that the PVC effects will be the same for the RMF and RHF calculations, and even for different Lagrangians.
Thus, the pseudodata shown here are mainly for the purpose of providing as much available information as possible, and they shall be treated only as a rough and supplementary reference.
The value of $\Delta \varepsilon \left( {Z = 58} \right)$ in the pseudodata is $0.76 \, \mathrm{MeV}$,
and the plot is moved downwards accordingly in Fig.~\ref{Fig:N82-pairing}(a).
The pseudodata also present a distinct kink, but the turning point is shifted to $Z = 58$.
Despite the change of the turning point in the pseudodata, they still decrease with the proton number first and then increase with it, which is the same as the case of the original data.
In particular, the slope of the original data is remarkably changed by the PVC effects only in the region between $Z = 58$ and $Z=64$, while it remains almost unchanged in other regions of both sides.
\par
The results calculated by the RHF theory using the density-dependent effective interactions PKA1~\cite{
  Long2007Phys.Rev.C76.034314}
and PKO$i$~($ i= 1 $, $ 2 $, $ 3 $)~\cite{
  Long2006Phys.Lett.B640.150,
  Long2008Europhys.Lett.82.12001}
are displayed in Fig.~\ref{Fig:N82-pairing}(a).
For the results of PKA1, PKO1, PKO2, and PKO3, the values of $\Delta \varepsilon \left( {Z = 58} \right)$ are $ 1.58 $, $ 2.84 $, $ 2.48 $, and $ 2.96 \, \mathrm{MeV} $, respectively.
It can be found that the energy differences calculated by PKO1 and PKO3 present distinct kinks at $Z = 58$, which is in consistence with the pseudodata.
In contrast, PKA1 gives a very small kink even without changing the sign of the slope, and PKO2 presents an almost linearly increasing trend.
\par
Using the method developed in Ref.~\cite{
  Wang2018Phys.Rev.C98.034313},
we calculate the contributions of the tensor force to the single-particle energies.
We then exclude the tensor-force contributions from the results of the full calculation.
The results without the contributions of the tensor force are shown in Fig.~\ref{Fig:N82-pairing}(b).
One can see that all the RHF effective interactions give almost identical results after excluding the tensor force.
This indicates that the differences in the results by different effective interactions mainly arise from the tensor forces.
In addition, the results without the tensor force present an approximately linearly increasing trend.
This shows that the nontrivial evolution of the single-particle energy differences is determined by the effects of the tensor force.
Through the characteristic spin dependence of the tensor force, one can further understand its effects on the above shell-structure evolution. According to the general understanding of the shell structure~\cite{
  Haxel1949Phys.Rev.75.1766,
  Mayer1949Phys.Rev.75.1969},
the single-particle level $\pi 1g_{7/2}$ ($\pi 1h_{11/2}$) is gradually occupied from $Z= 50$
($Z= 64$).
The tensor-force effect between $\pi 1g_{7/2}$ and $\nu 1i_{13/2}$ is attractive, while it turns to be repulsive between $\pi 1g_{7/2}$ and $\nu 1h_{9/2}$.
Consequently, the tensor-force effects arising from the occupation of $\pi 1g_{7/2}$ reduce the energy difference between the states $\nu 1i_{13/2}$ and $\nu 1h_{9/2}$.
Due to the similar mechanism, the tensor-force effects arising from the occupation of $\pi 1h_{11/2}$ increase the energy difference under discussion.
\par
In Ref.~\cite{
  Wang2018Phys.Rev.C98.034313},
it was demonstrated that the tensor force arising from the $\pi$-PV coupling dominates over those from all the other nucleon-meson couplings.
This is true not only for PKA1 but also for PKO1 and PKO3.
As shown in Eq.~\eqref{Eq:DDPI}, the $\pi$-PV coupling strength in the effective interactions adopted here depends on the baryon density exponentially.
Its value at zero density, i.e., $f_{\pi} \left( 0 \right)$,
in PKO3 is the same as that in PKO1, but the factor of density dependence $a_{\pi}$ in PKO3 is smaller~\cite{
  Long2008Europhys.Lett.82.12001}.
Therefore, the tensor force in PKO3 is stronger than that in PKO1.
In PKA1, the contribution of the tensor force from the $\rho$-T coupling, which is not considered in the PKO series, remarkably cancels that from the $\pi$-PV one~\cite{
  Wang2018Phys.Rev.C98.034313}.
As a consequence, the net tensor-force effect in PKA1 is weaker than those in both PKO1 and PKO3.
Notice that the $\pi$-PV coupling is not included in PKO2 at all.
According to the discussion above, one can recognize that, for the strength of the tensor force,
$ \text{PKO3} > \text{PKO1} > \text{PKA1} > \text{PKO2} $.
Keeping this conclusion in mind, let us turn back to Fig.~\ref{Fig:N82-pairing}(a) again.
The following two criteria are adopted to evaluate the theoretical description of the shell-structure evolution here:
(i) whether it can give a turning or kink;
and (ii) how well the slope reproduces the (pseudo)data.
The position of the turning point is deeply related to the shell structures at $Z = 58$ and $Z= 64$.
In fact, the artificial shell at $Z = 58$ and the subshell at $Z=64$ have been extensively studied within the framework of CDFT~\cite{
  Long2007Phys.Rev.C76.034314,
  Geng2019Phys.Rev.C100.051301,
  Wei2020Chin.Phys.C44.074107}.
With the criteria above, one can see that PKO3 reproduces the data best among all the RHF effective interactions.
While it is not as good as PKO3, PKO1 is much better than PKA1 and PKO2.
Considering the relative strengths of the tensor forces in these RHF effective interactions, one can conclude that the effective interaction with stronger tensor force reproduces the evolution of the single-particle energy difference better.
Further, it implies that the tensor force in the current RHF effective interactions seems in general too weak, and stronger tensor force is appealing for the future development of the effective interactions.
\begin{figure*}[tb]
  \includegraphics[width=0.45\textwidth]{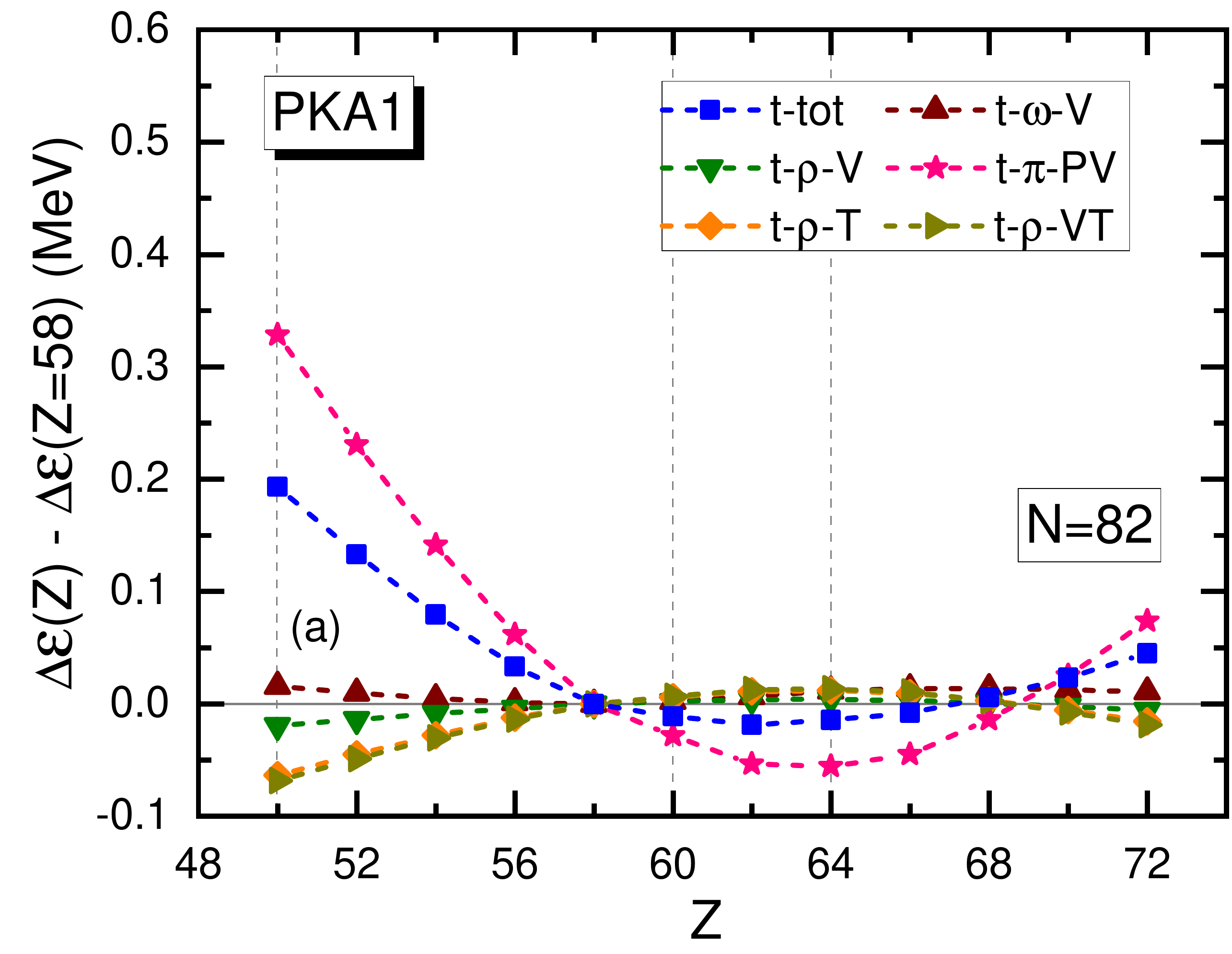}
  \hspace{0.5cm}
  \includegraphics[width=0.45\textwidth]{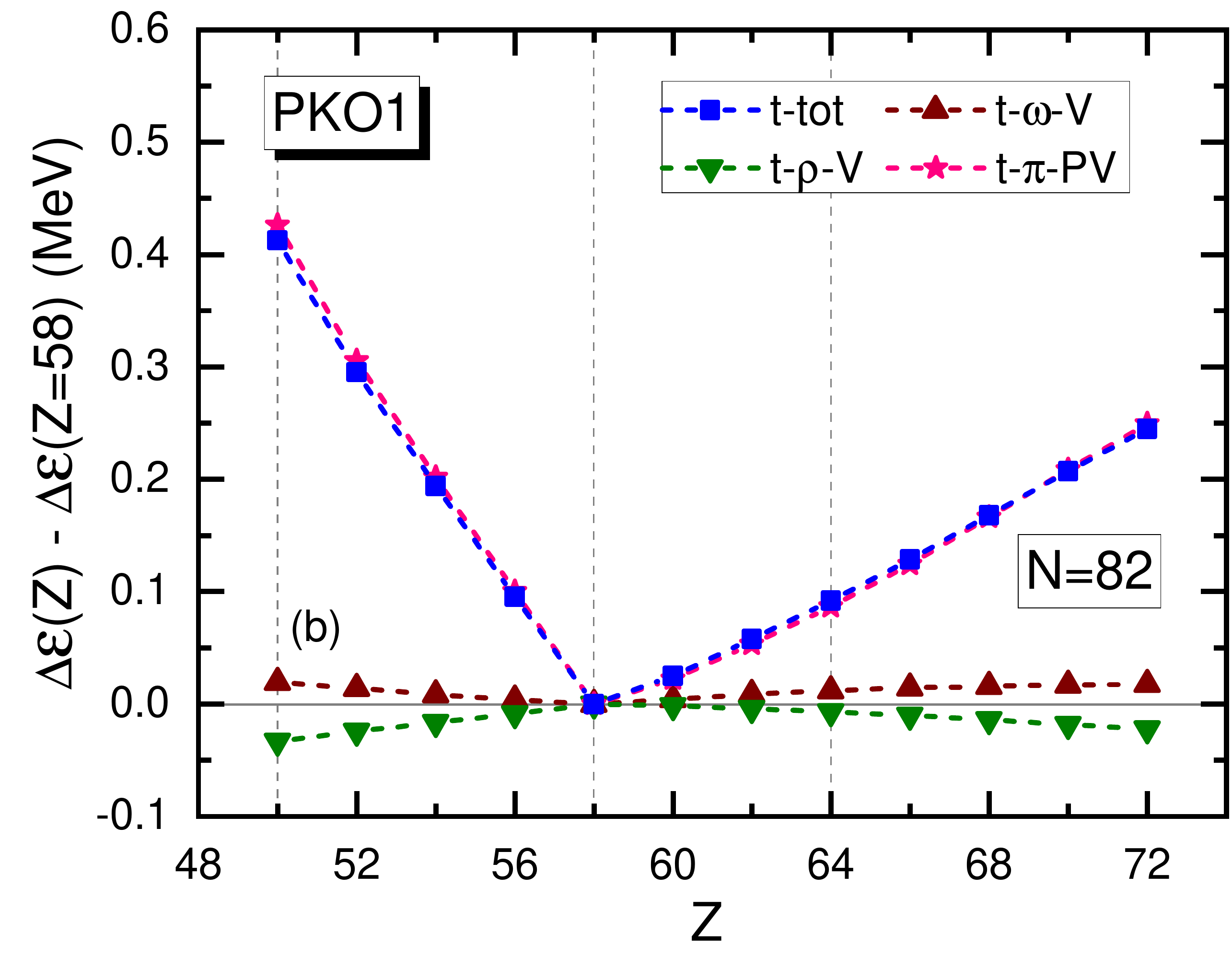}
  \vspace{0.5cm}
  \includegraphics[width=0.45\textwidth]{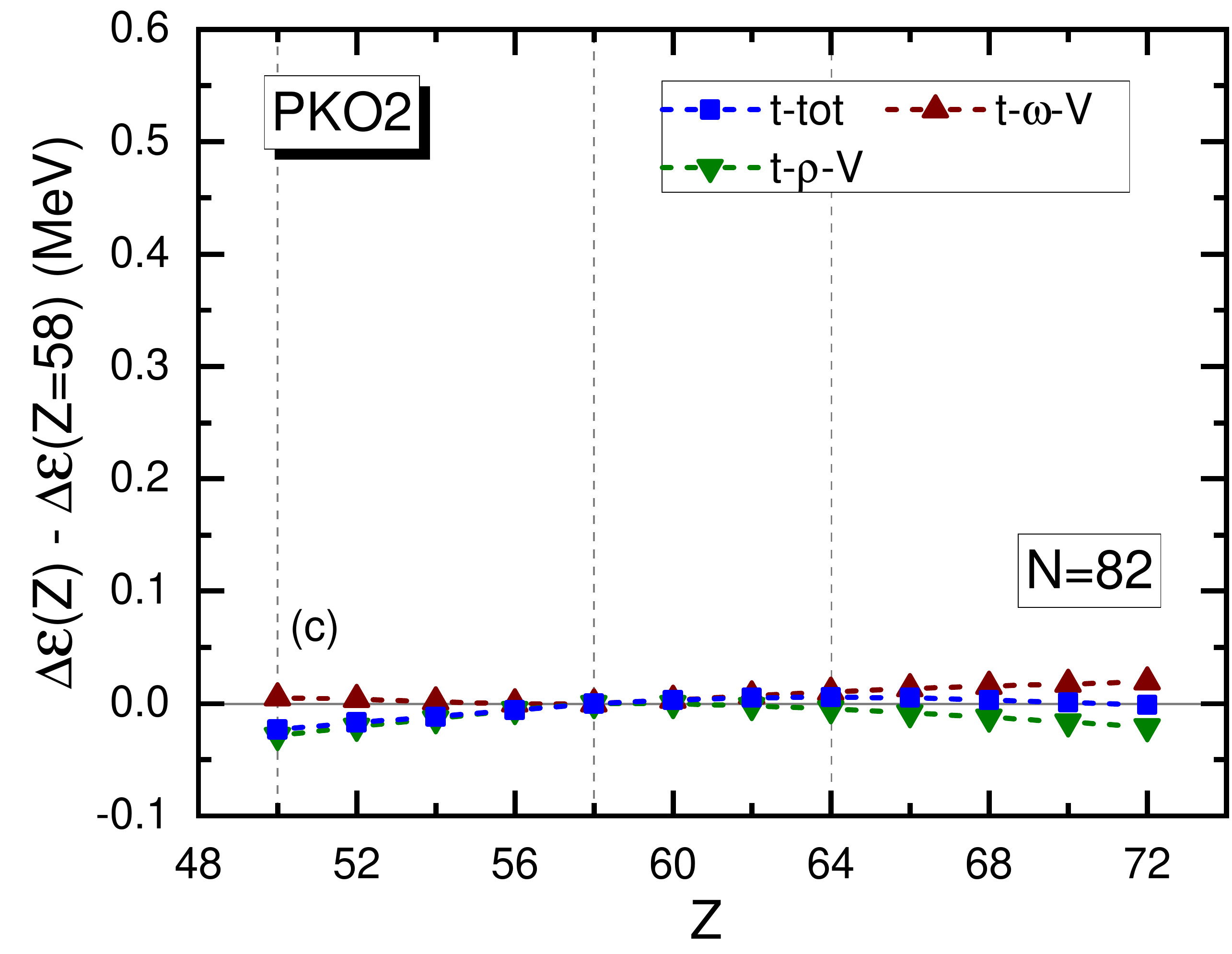}
  \hspace{0.5cm}
  \includegraphics[width=0.45\textwidth]{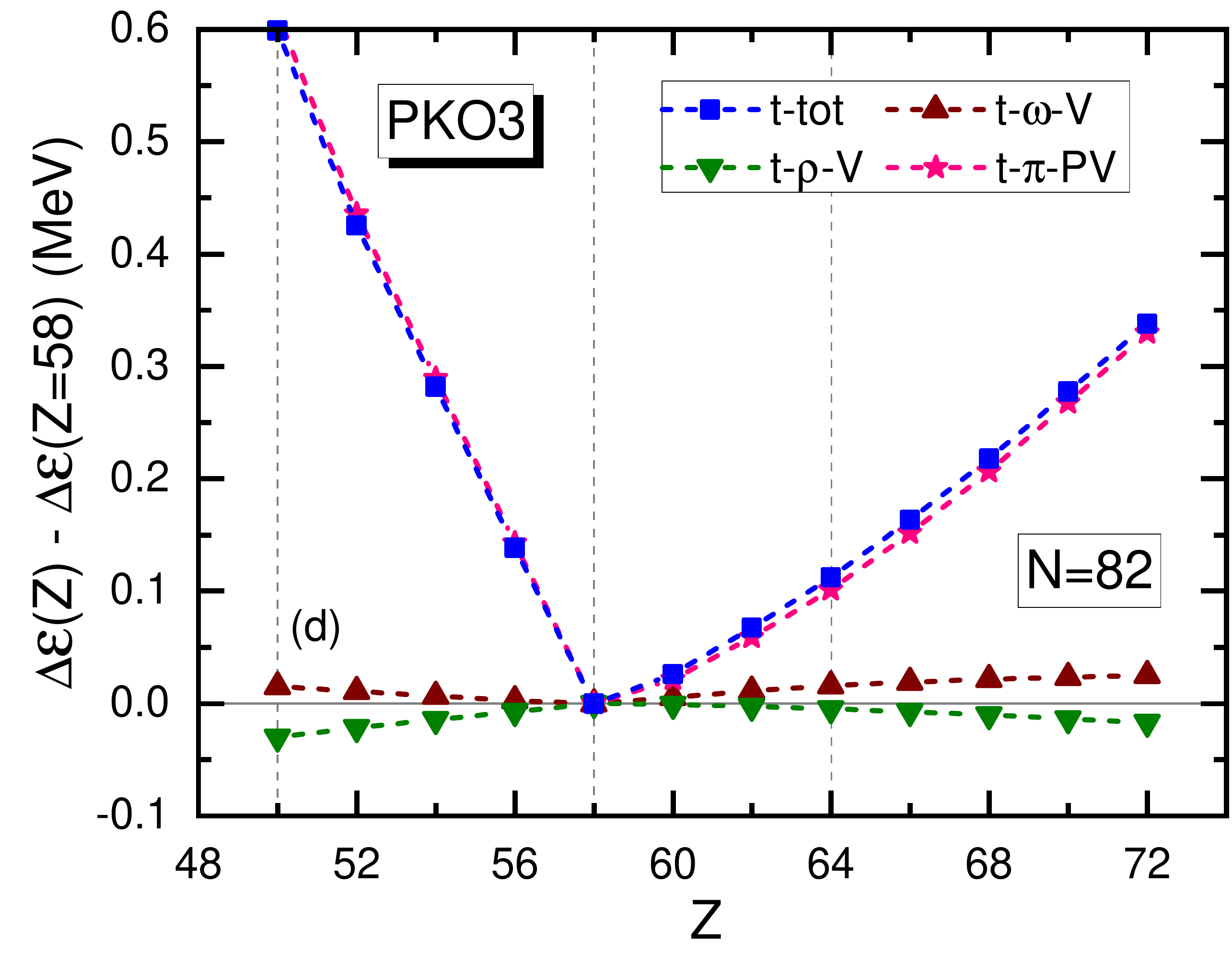}
  \caption{
    The tensor-force contributions to the energy differences
    $\Delta \varepsilon \equiv \varepsilon_{\nu 1i_{13/2}}-\varepsilon_{\nu 1h_{9/2}}$
    in the $N=82$ isotones as functions of the proton number, calculated by the RHF theory with the effective interactions PKA1~\cite{Long2007Phys.Rev.C76.034314} and PKO$i$ ($ i = 1 $, $ 2 $, $ 3 $)~\cite{Long2006Phys.Lett.B640.150, Long2008Europhys.Lett.82.12001}.
    The blue filled squares denote the total contributions of the tensor force, while the contributions of the individual coupling are denoted by the other symbols.
    All the results are normalized with respect to their corresponding values at $Z = 58$.}
  \label{Fig:N82-tensor-force}
\end{figure*}
\par
Next, we will discuss explicitly the effects of the tensor force on the evolution of the energy difference between the states $\nu 1i_{13/2}$ and $\nu 1h_{9/2}$ along the $N=82$ isotones.
The contributions of the tensor force to the energy differences, calculated with the effective interactions PKA1 and the PKO series, are shown in the different panels of Fig.~\ref{Fig:N82-tensor-force}.
\par
For the results of PKA1, shown in Fig.~\ref{Fig:N82-tensor-force}(a), the net contribution of the tensor forces of all the involved nucleon-meson couplings has a flat minimum around $Z = 64$, which is at same position of the minimum of the original experimental data.
As expected, the tensor force from the $\pi$-PV coupling is the most remarkable compared with those from the other couplings, and it also gives a minimum around $Z = 64$.
The trend of the net contribution of the tensor forces is mainly determined by the tensor force from the $\pi$-PV coupling.
Meanwhile, one can see that the contribution of the tensor force from the $\pi$-PV coupling is partially canceled by those from the other couplings, especially the $\rho$-T and $\rho$-VT ones.
Since the tensor-force contribution from the $\omega$-V coupling comes from the neutron-neutron interaction, rather than the proton-neutron one, it is not so relevant for the shell-structure evolution discussed here.
\par
From Figs.~\ref{Fig:N82-tensor-force}(b) and~\ref{Fig:N82-tensor-force}(d),
it can be seen that the net contributions of the tensor force calculated by PKO1 and PKO3 present quite sharp turnings at $Z = 58$, which is exactly the turning point of the pseudodata.
Based on the comparison of the results with and without the tensor force displayed in Fig.~\ref{Fig:N82-pairing},
one can conclude that the reason why PKO1 and PKO3 can present distinct kinks at $Z = 58$ is largely related to the sharp turning of the tensor-force contribution.
Moreover, it is noticeable  that the net tensor-force contributions in both PKO1 and PKO3 are almost fully determined by the contribution of the $\pi$-PV coupling.
This is because the tensor force from the $\omega$-V coupling largely cancels that from the $\rho$-V one.
Here, we remind again that the effects of the tensor force from the $\omega$-V coupling are due to the neutron-neutron rather than the proton-neutron interactions.
Such a cancellation does not appear for PKA1, mainly due to the considerable contributions of the tensor force from $\rho$-T coupling as well as the $\rho$-VT one.
In addition, one finds that the contribution of the tensor force in PKO3 is larger than that in PKO1.
This results from the fact that the coupling strength of $\pi$-PV coupling in PKO3 is larger, as mentioned above.
The results of PKO2 are very different.
Because of the absence of both $\pi$-PV and $\rho$-T couplings~\cite{
  Long2008Europhys.Lett.82.12001},
the proton-neutron tensor force in PKO2 is very weak.
Therefore, the net contribution of the tensor force in PKO2 is almost negligible for the shell-structure evolution discussed here, as can be seen from Fig.~\ref{Fig:N82-tensor-force}(c).
\subsection{$Z = 50$ isotopes}\label{Subsec:Z50}
\begin{figure*}[tb]
  \includegraphics[width=0.45\textwidth]{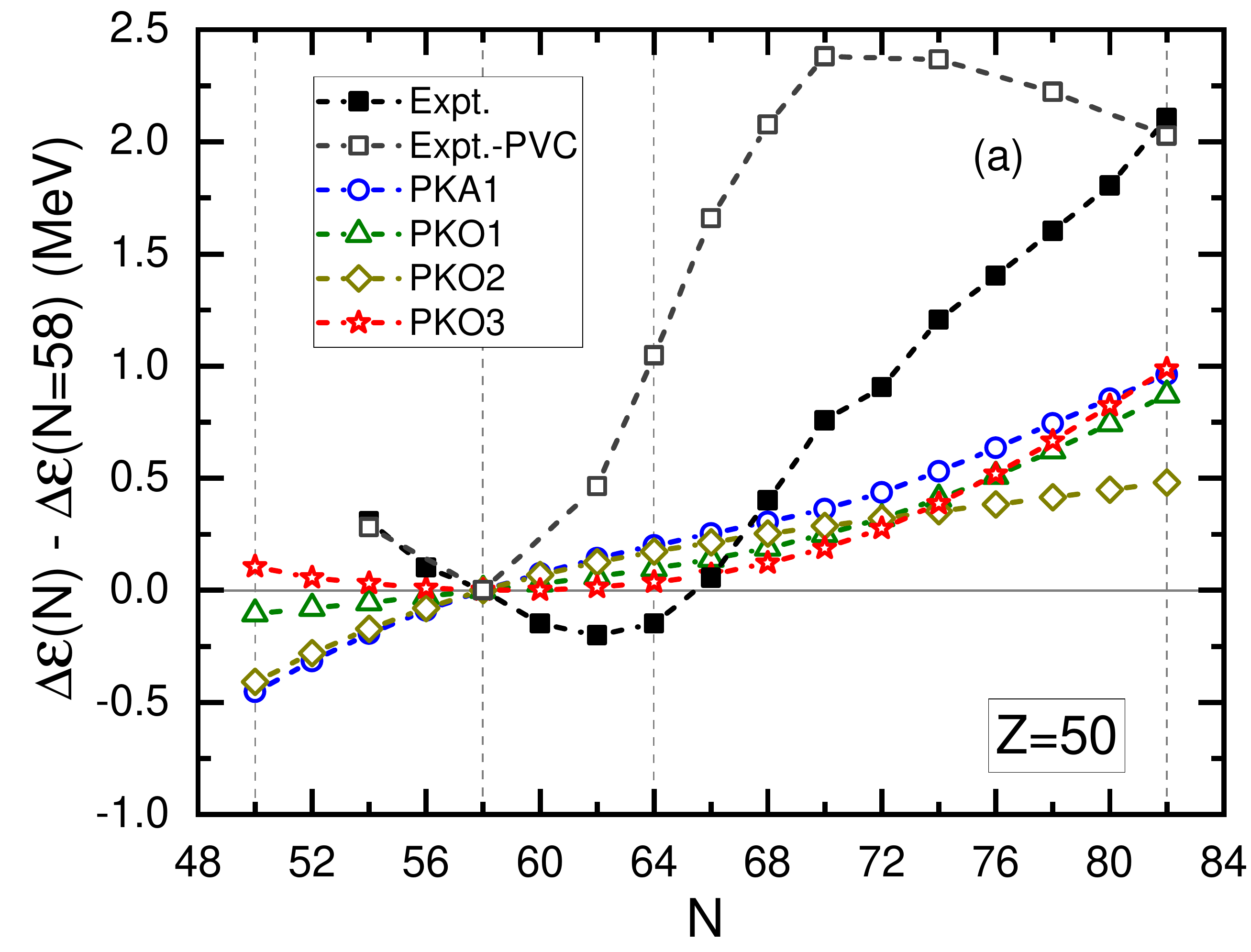}
  \hspace{0.5cm}
  \includegraphics[width=0.45\textwidth]{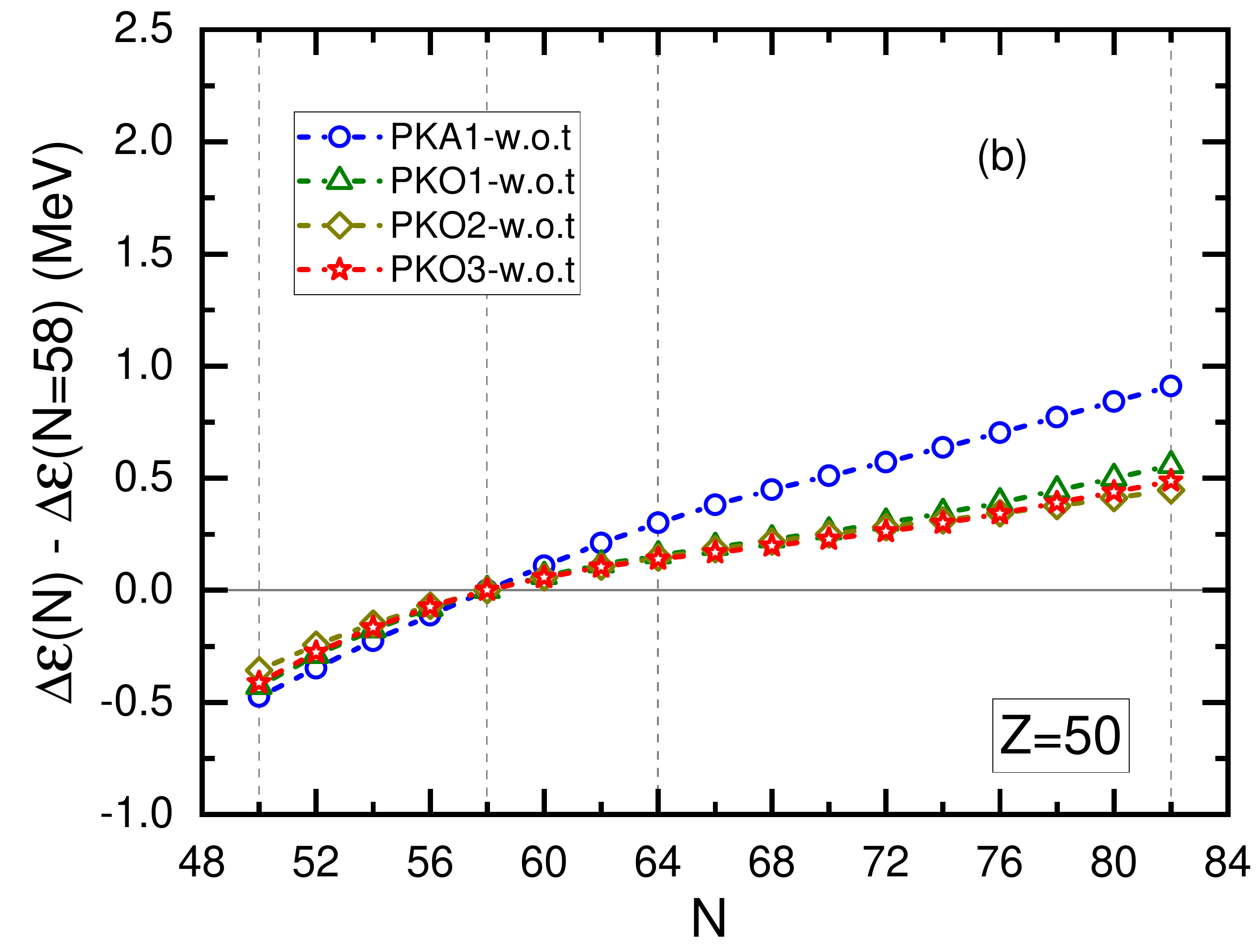}
  \caption{
    Similar to Fig.~\ref{Fig:N82-pairing}, but for the energy differences
    $\Delta \varepsilon \equiv \varepsilon_{\pi 1h_{11/2}}-\varepsilon_{\pi 1g_{7/2}}$
    in the $Z=50$ isotopes as functions of the neutron number.
    All the experimental data and the calculated results are normalized with respect to their corresponding values at $N = 58$.}
  \label{Fig:Z50-pairing}
\end{figure*}
\par
Figure~\ref{Fig:Z50-pairing} displays the energy differences between the single-particle states
$\pi 1h_{11/2}$ and $\pi 1g_{7/2}$ in the $Z = 50$ ($ \mathrm{Sn} $) isotopes as functions of the neutron number.
Similar with the case of the $N = 82$ isotones studied above, Fig.~\ref{Fig:Z50-pairing}(a) displays both the original experimental data and the pseudodata, where the latter are obtained in the same way as those in the $N = 82$ isotones.
The calculations are also performed in the same way as those for the $N = 82$ isotones.
All the experimental data and the calculated results are normalized with respect to their corresponding values at $N = 58$.
The values of $\Delta \varepsilon \left( {N = 58} \right)$ in the original data and pseudodata are $ 0.69 $ and $ 1.58 \, \mathrm{MeV} $, respectively.
One can find that the original data present a minimum at $N = 62$, while it is shifted to $N = 58$ after considering the PVC effects based on RMF.
Meanwhile, the PVC effects do not modify remarkably the slope in the region from $N = 54$ to $N = 58$ and that in the region from $N = 64$ to $N = 70$.
\par
For the results of PKA1, PKO1, PKO2, and PKO3, the values of $\Delta \varepsilon \left( {N = 58} \right) $ are $ 3.05 $, $ 4.32 $, $ 3.63 $, and $ 4.54 \, \mathrm{MeV} $, respectively.
Among all the RHF effective interactions used here, only PKO3 gives a visible minimum, which is located around $N = 58$.
Even though the slope near the minimum is quite flat, it qualitatively reproduces the trend given by the (pseudo)data, i.e., the energy difference decreases first and then increases with the neutron number.
In contrast, the energy differences calculated by all other effective interactions increase monotonically with the neutron number, failing to reproduce the experimental trend even qualitatively.
Compared with PKA1 and PKO2, the results of PKO1 are relatively closer to those of PKO3, also closer to the data.
Given the relative strengths of the tensor forces in these effective interactions, i.e.,
$ \text{PKO3} > \text{PKO1} > \text{PKA1} > \text{PKO2} $,
one can suppose again that the tensor force is the main cause of the differences among these results.
To make it clearer, we subtract the contributions of the tensor force following what has been done for the $N = 82$ isotones. The results are displayed in Fig.~\ref{Fig:Z50-pairing}(b).
Without the contributions of the tensor force, all the RHF effective interactions give similar results, i.e., the energy differences increase monotonically with the neutron number.
Thus, the determinant role of the tensor force is confirmed again.
\begin{figure*}[tb]
  \includegraphics[width=0.45\textwidth]{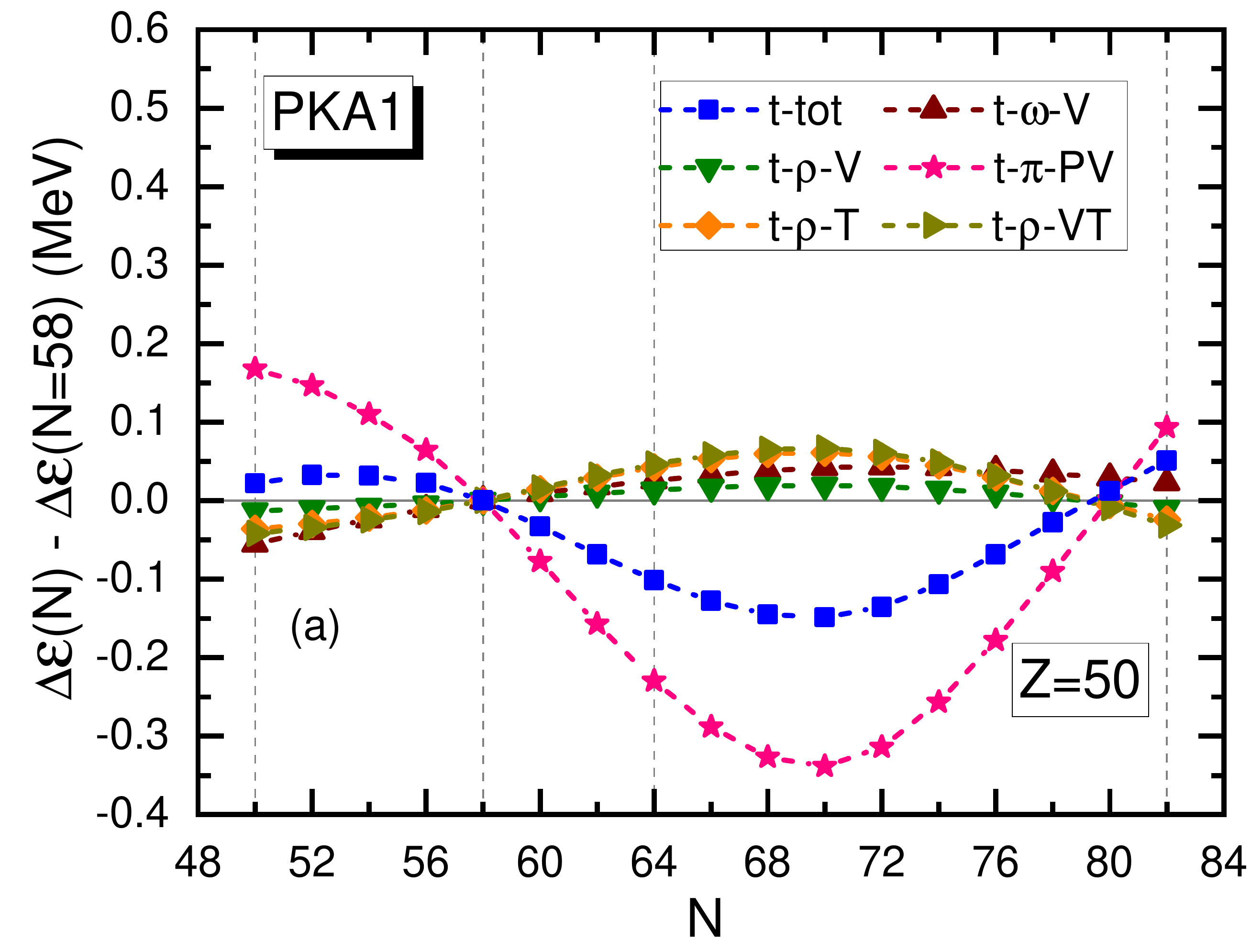}
  \hspace{0.5cm}
  \includegraphics[width=0.45\textwidth]{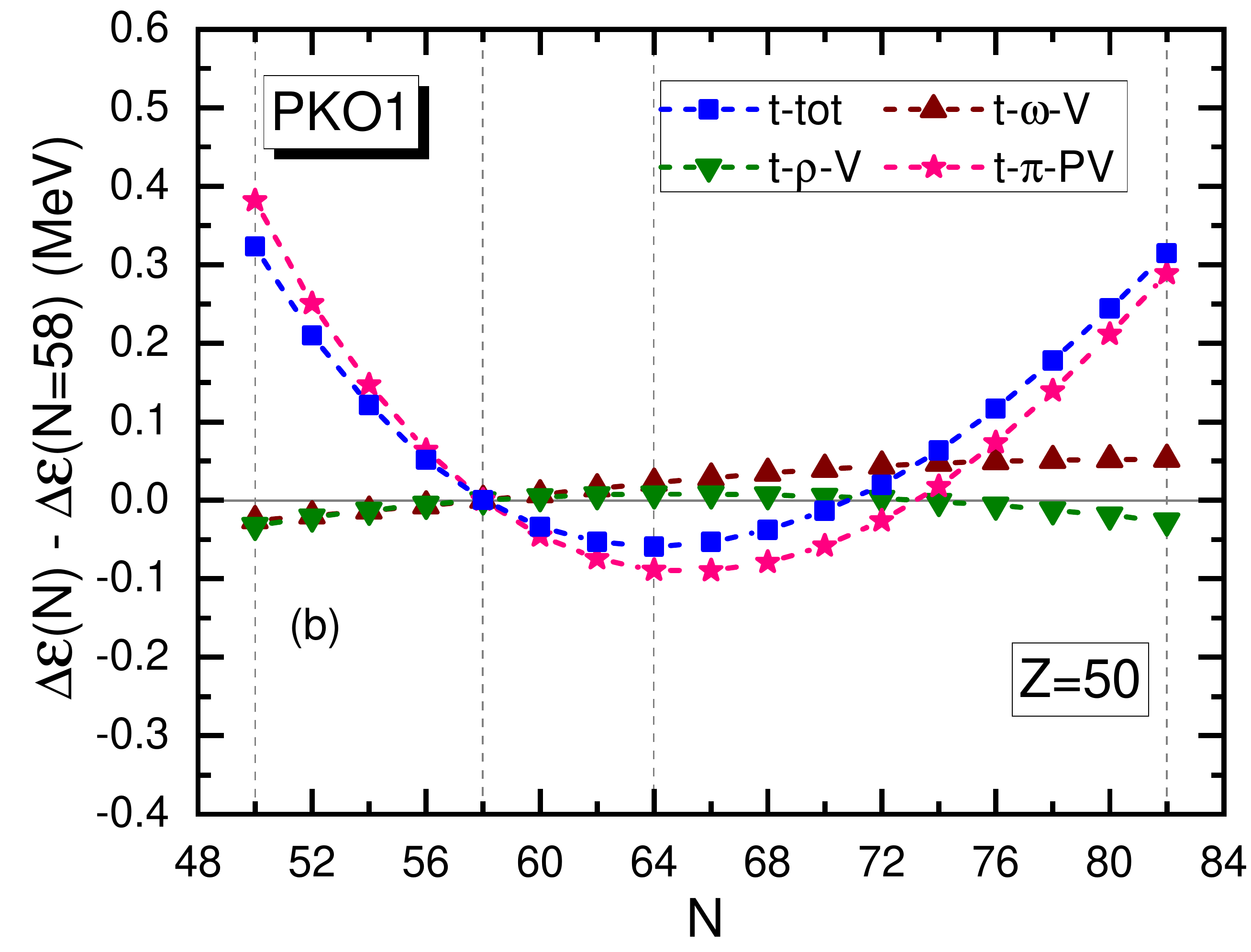}
  \vspace{0.5cm}
  \includegraphics[width=0.45\textwidth]{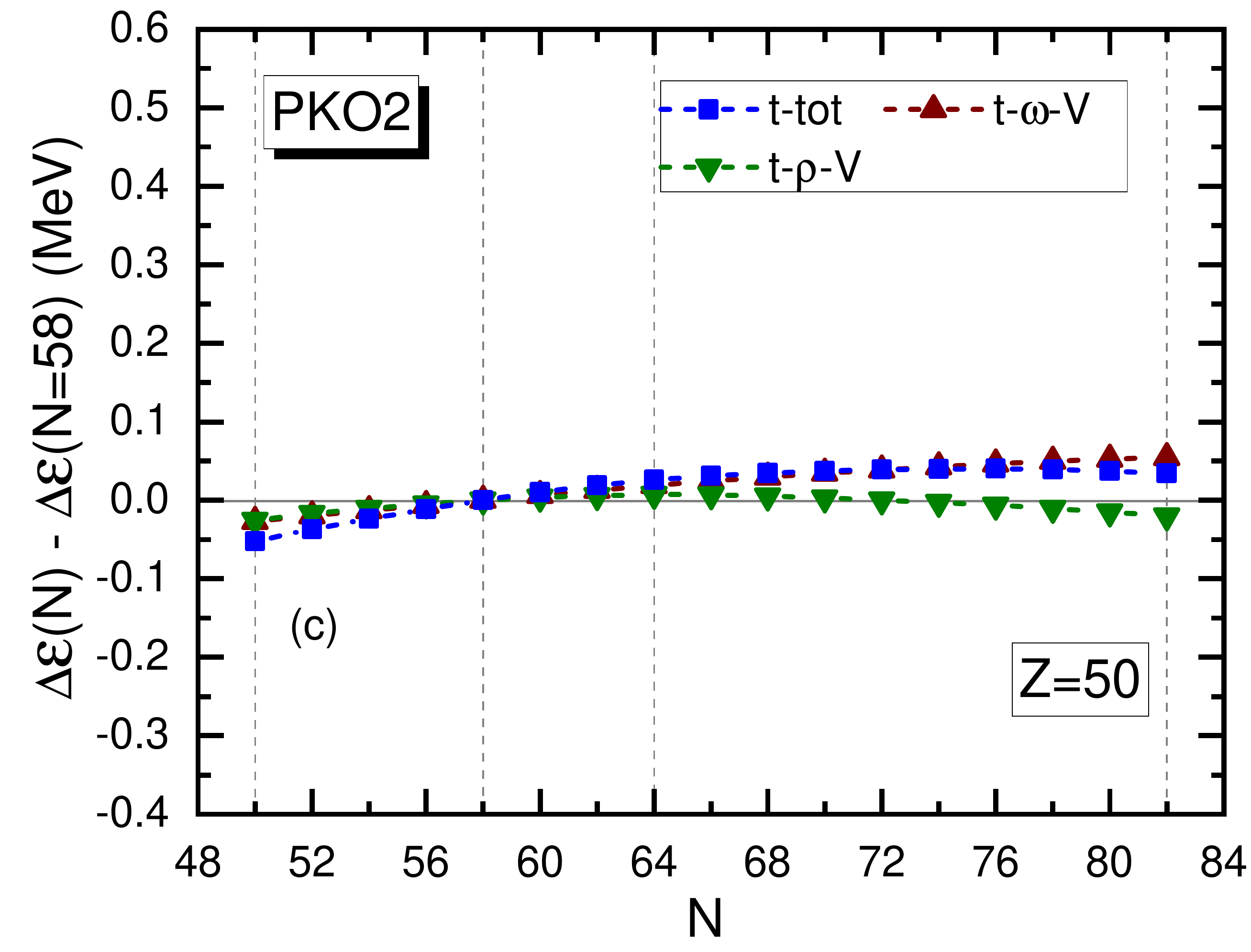}
  \hspace{0.5cm}
  \includegraphics[width=0.45\textwidth]{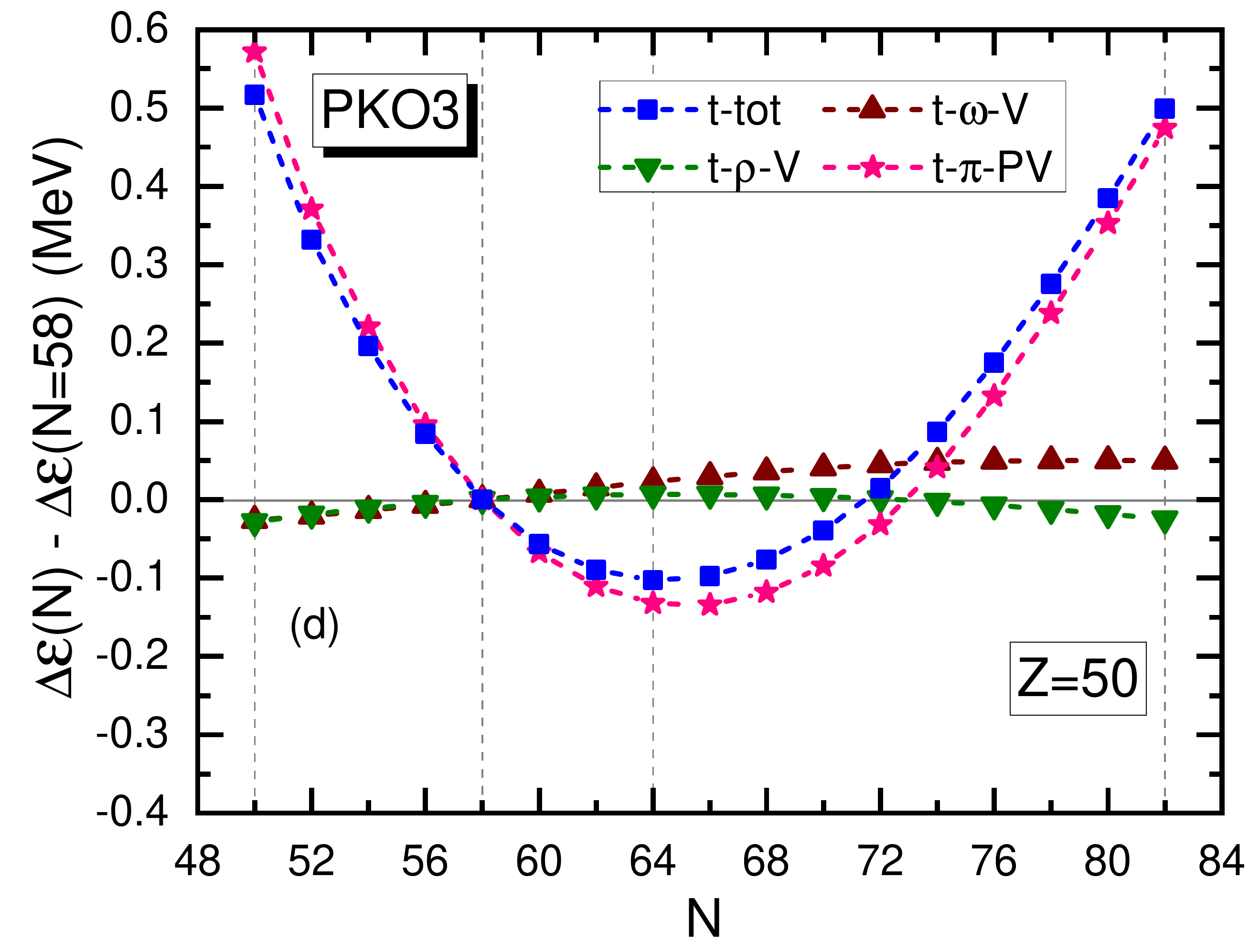}
  \caption{
    Similar to Fig.~\ref{Fig:N82-tensor-force}, but for the energy differences
    $\Delta \varepsilon \equiv \varepsilon_{\pi 1h_{11/2}}-\varepsilon_{\pi 1g_{7/2}}$ in the $Z= 50$ isotopes as functions of the neutron number. All the results are normalized with respect to their corresponding values at $N = 58$.}
  \label{Fig:Z50-tensor-force}
\end{figure*}
\par
Shown in Fig.~\ref{Fig:Z50-tensor-force} are the explicit contributions of the tensor force calculated by PKA1 and PKO series.
The net contribution of the tensor force in PKA1, shown in Fig.~\ref{Fig:Z50-tensor-force}(a), has a minimum around $N = 70$, which is consistent with neither the original data nor the pseudodata.
The contribution of the tensor force from the $\pi$-PV coupling dominates over those from the other couplings.
The net contribution of the tensor force is qualitatively in line with that from the $\pi$-PV coupling, and the former is smaller than the latter in amplitude.
This is determined by the relative strength of the tensor force in each nucleon-meson coupling and its sign.
The feature is the same as that in the case of $N = 82$ isotones. The details will not be repeated here.
\par
The contributions of the tensor force in PKO1 and PKO3 are shown in Figs.~\ref{Fig:Z50-tensor-force}(b) and~\ref{Fig:Z50-tensor-force}(d), respectively.
It can be seen that the minima given by PKO1 and PKO3 are both near $N = 64$, which is the same position of the minimum of the original data shown in Fig.~\ref{Fig:Z50-pairing}.
As mentioned above, the minimum of the energy differences calculated by PKO3, which is shown in Fig.~\ref{Fig:Z50-pairing}, is near $N = 58$.
This means the minimum of the contribution of tensor force cannot uniquely determine the minimum of the energy difference.
Even though the contributions of the tensor forces given by PKO1 and PKO3 are similar to each other, the former fails to reproduce the experimental trend of the shell-structure evolution.
Similar to the results for the $N = 82$ isotones, the contribution of the tensor force in PKO3 is larger than that in PKO1, as can be seen in Figs.~\ref{Fig:Z50-tensor-force}(b) and~ \ref{Fig:Z50-tensor-force}(d).
This reminds us that the strength of the tensor force is crucial for reproducing the trend of the experimental data.
For PKO2, the contribution of the tensor force is minor due to the absence of the $\pi$-PV coupling, as shown in Fig.~\ref{Fig:Z50-tensor-force}(c).
\par
It is interesting to see the differences between the contributions of the tensor force in the $N = 82$ isotones and those in the $Z = 50$ isotopes.
For the calculation with PKO1 and PKO2, the contribution of the tensor force in the $Z = 50$ isotopes changes gently around the turning point, while the turning at $Z = 58$ in the $N = 82$ isotones is quite sharp. Such a difference can be understood through the effects of pairing.
For the $N = 82$ isotones, since PKO1 and PKO3 give large artificial shells at $Z = 58$, the protons occupy gradually the orbital $\pi 1g_{7/2}$ from $Z = 50$ to $Z = 58$, leaving the orbital $\pi 2d_{5/2}$ almost empty.
The orbitals $\pi 1g_{7/2}$ and $\pi 2d_{5/2}$ are spin down and spin up, respectively, and their tensor-force effects on the energy differences under discussion here are opposite to each other.
It is the sudden change of the occupation of $\pi 2d_{5/2}$ at $Z = 60$ that results in the sharp turning.
In contrast, the $N = 58$ shell gaps in the $Z = 50$ isotopes are not so pronounced in the calculations here.
Therefore, for the $Z = 50$ isotopes with the neutron numbers from $52$ to $64$,
both the orbitals $\nu 1g_{7/2}$ and $\nu 2d_{5/2}$ have nonvanishing occupation probabilities and their tensor-force effects partially cancel each other.
This is the key reason why there is no sharp change around $N = 58$.
The properties of the slope around $ N $, $ Z = 64 $ and $70$, which belong to the critical subshells, can also be understood through the magnitudes of the relevant gaps and the pairing effects.
We will no longer go into the details since the mechanism is similar.
\subsection{Exploration of the strength of tensor force}
\par
Based on the discussion above, one finds that the tensor force in the current RHF effective interactions may be too weak to reproduce the evolution of the single-particle energy differences.
In fact, such a point of view was declared more than ten years ago~\cite{
  Long2008Europhys.Lett.82.12001}.
A similar conclusion was also drawn through the analysis of the evolutions of several magical shells~\cite{
  Wang2018Phys.Rev.C98.034313}
and  the spin-orbit splittings in the neutron drops~\cite{
  Shen2017Phys.Rev.C96.014316,
  Shen2018Phys.Lett.B778.344}.
Considering that the $\pi$-PV coupling is the most important carrier of the tensor force, it is natural to explore the strength of the tensor force by enlarging the $\pi$-PV coupling strength $f_{\pi}$.
Through such an investigation, we expect to give some guidance for developing the effective interactions with well constrained tensor forces in the relativistic framework.
\par
Considering the exponential density dependence of $f_{\pi}$, as shown in Eq.~\eqref{Eq:DDPI},
one can enlarge it in two ways:
(i)~multiplying a factor $\lambda$ ($\lambda > 1$) as a whole,
or (ii)~weakening the density dependence by multiplying a factor $\eta$ ($\eta < 1$) in the coefficient $a_{\pi}$.
Obviously, the latter strategy does not change the value at zero density, but makes $f_{\pi}$ decrease more slowly with the baryon density.
\begin{figure*}
  \includegraphics[width=0.45\textwidth]{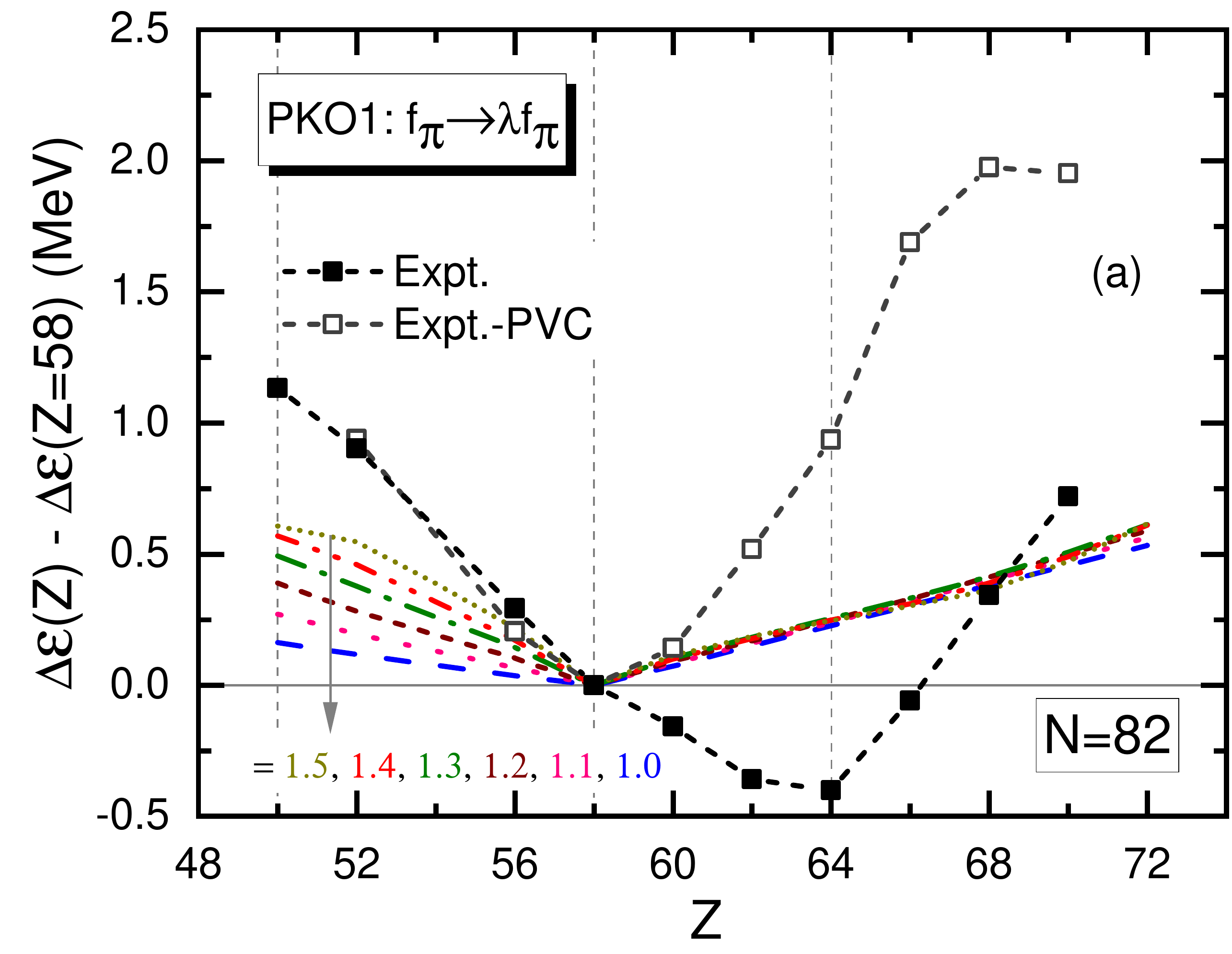}
  \hspace{0.5cm}
  \includegraphics[width=0.45\textwidth]{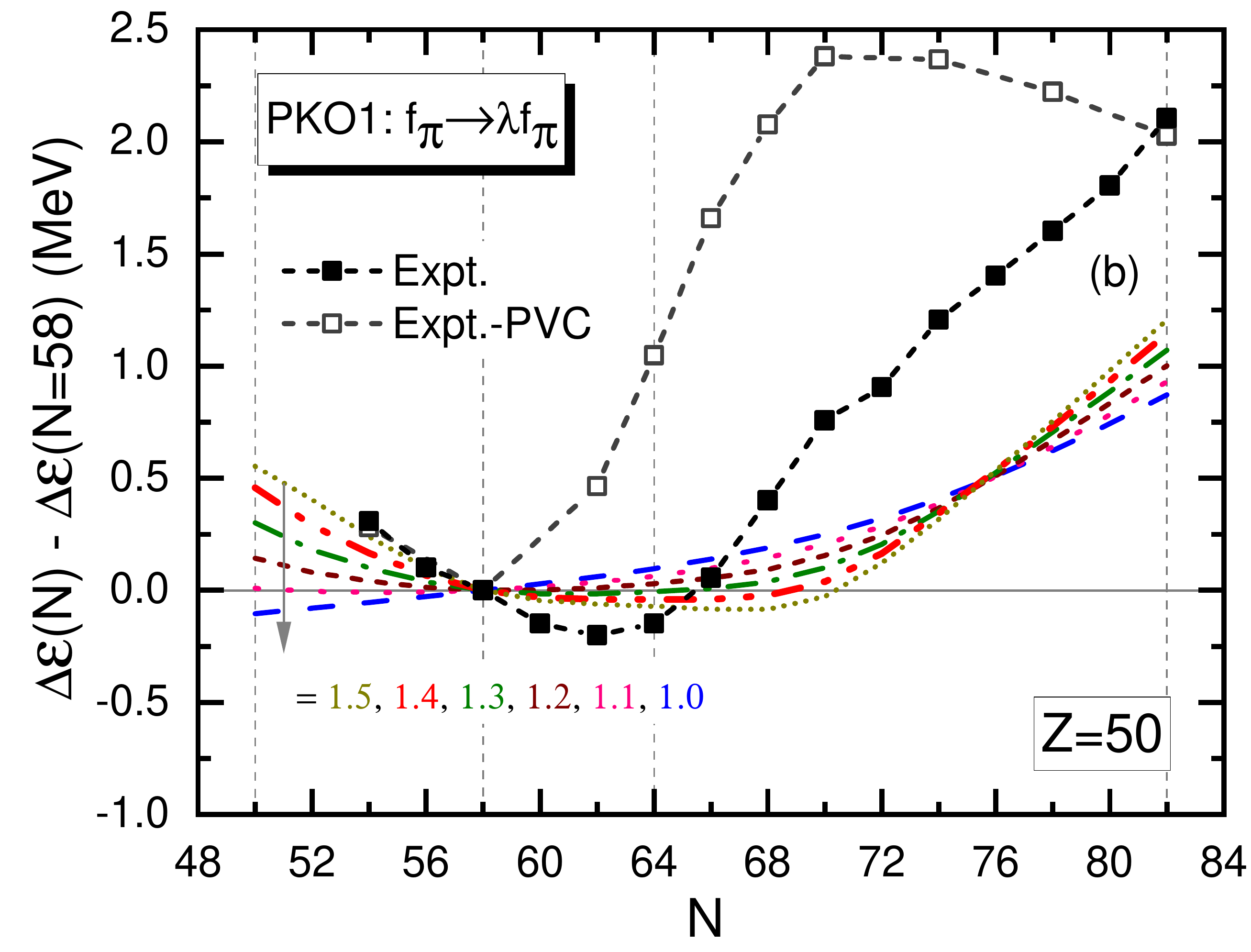}
  \caption{
    (a) Energy differences
    $\Delta \varepsilon \equiv \varepsilon_{\nu 1i_{13/2}}-\varepsilon_{\nu 1h_{9/2}}$
    in the $N=82$ isotones as functions of the proton number,
    normalized with respect to the values at $Z = 58$.
    (b) Energy differences
    $\Delta \varepsilon \equiv \varepsilon_{\pi 1h_{11/2}}-\varepsilon_{\pi 1g_{7/2}}$
    in the $Z=50$ isotopes as functions of the neutron number,
    normalized with respect to the values at $N = 58$.
    The calculation is performed by the RHF theory with the effective interaction PKO1,
    but the $\pi$-PV coupling is enlarged by multiplying a factor $\lambda$, i.e., $f_{\pi} \rightarrow \lambda f_{\pi}$
    ($\lambda = 1.0 $, $ 1.1 $, $ 1.2 $, $ 1.3 $, $ 1.4 $, and $1.5$).
    For comparison, the experimental data are also given, which are the same as those in Figs.~\ref{Fig:N82-pairing} and \ref{Fig:Z50-pairing}.}
  \label{Fig:xfpi}
\end{figure*}
\par
It is noticeable that PKO1 is the first widely used RHF effective interaction.
More importantly, it qualitatively reproduces the shell-structure evolution due to the tensor force mainly from the $\pi$-PV coupling, and so does PKO3, as discussed in Secs.~\ref{Subsec:N82} and \ref{Subsec:Z50}.
In contrast, PKO2 cannot reproduce the shell structure for the absence of the $\pi$-PV coupling; PKA1 also fails, because the tensor forces from the $\rho$-T and $\rho$-VT couplings partially cancel that from the $\pi$-PV coupling.
For a primary study of the strength of the tensor force, it is efficient to focus on the dominant carrier of the tensor force, i.e., the $\pi$-PV coupling, and avoid the possible distraction of the cancellation from $\rho$-T and $\rho$-VT couplings.
Therefore, we take PKO1 as an example and investigate how the enhancement of $f_{\pi}$, in the two different ways above, affects the description of the evolution of the energy difference discussed in the previous subsections.
First, we multiply $f_{\pi}$ by different $\lambda$
($\lambda = 1.0 $, $ 1.1 $, $ 1.2 $, $ 1.3 $, $ 1.4 $, and $1.5$),
and then calculate the energy differences between the states $\nu 1i_{13/2}$ and $\nu 1h_{9/2}$ in the $N=82$ isotones and those between the states $\pi 1h_{11/2}$ and $\pi 1g_{7/2}$ in the $Z = 50$ isotopes.
In Fig.~\ref{Fig:xfpi}(a), it can be seen that with the enhancing of the $\pi$-PV coupling, the results of $N = 82$ isotones change remarkably and approach the data gradually in the region of $Z \leq58 $.
Similar properties are found for the $Z = 50$ isotopes in the region of $N \leq 58 $.
Note that in the regions of $ N $, $ Z > 58 $, the energy differences in both the isotonic and isotopic chains are not so sensitive to the multiplier $\lambda$.
In particular, for the $N=82$ isotones, the energy differences are almost independent to $\lambda$ when $Z > 58$.
It is worth mentioning once more that for the $N = 82$ isotones ($Z = 50$ isotopes) with
$ \text{$ Z $~($ N $)} \leq 58 $ and $ 64 \leq \text{$ Z $~($ N $)} \leq 68 $, the original data and the pseudodata present almost the same slopes.
Thus, the data in these regions are supposed to be more informative for the mean-field calculations.
When $\lambda = 1.5$, the data of the $Z = 50$ isotopes with $N \leq 58$ are reproduced quite well.
However, for the $N = 82$ isotones with $Z \leq 58$, it seems that one needs to further enlarge the $f_{\pi}$.
Nevertheless, this may not work so well.
One can see that the trend starts to become flat around $Z = 50$ when $\lambda = 1.5$.
This means that when $\lambda$ is too large, the shell closure at $Z = 50$ breaks down and the pairing correlation arises.
Therefore, it is unlikely to reproduce the data better with even larger $\lambda$.
Meanwhile, larger $\lambda$ may worsen the description of the $Z = 50$ isotopes with overestimated slope in the region of $N \leq 58$.
\par
\begin{figure*}
  \includegraphics[width=0.45\textwidth]{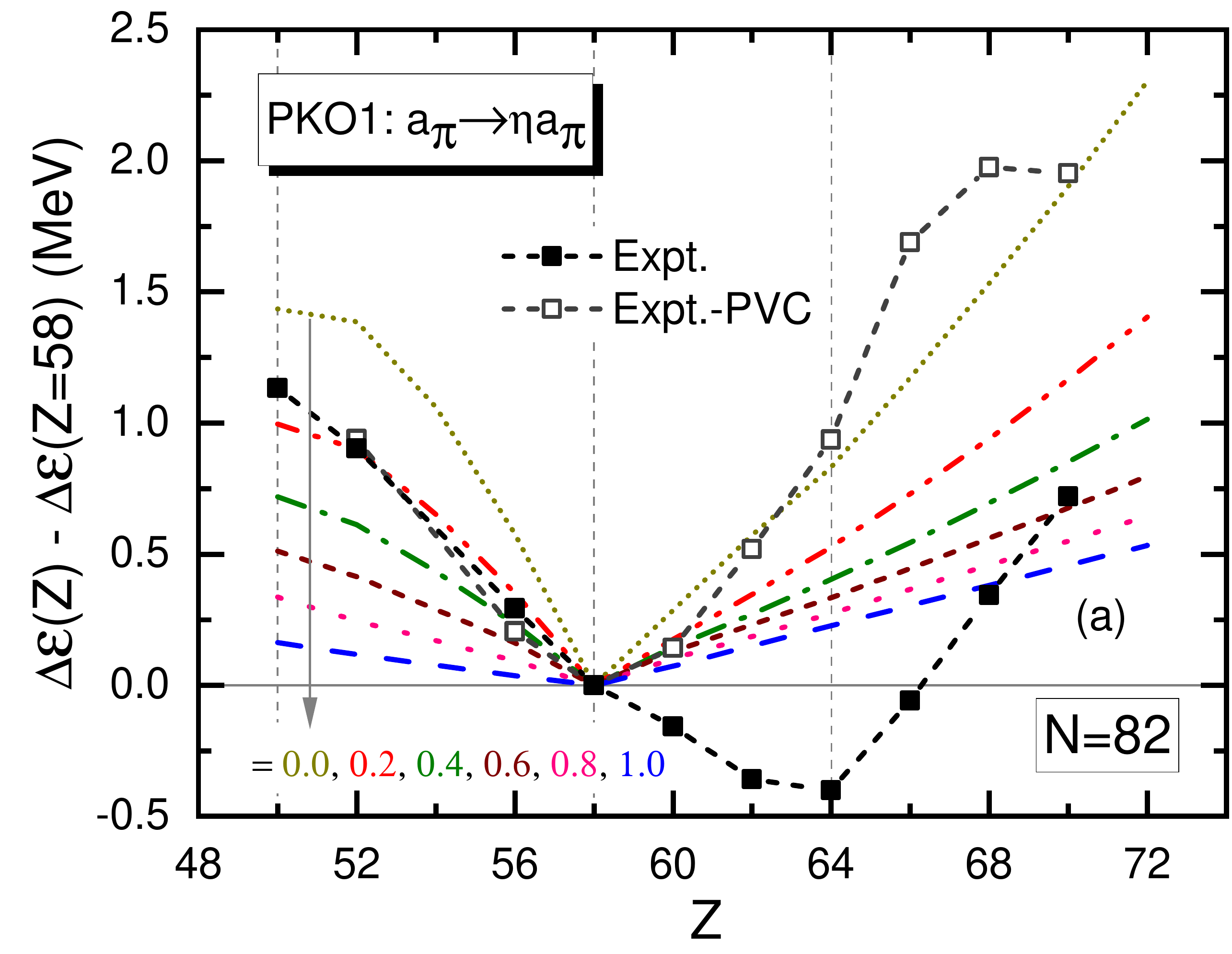}
  \hspace{0.5cm}
  \includegraphics[width=0.45\textwidth]{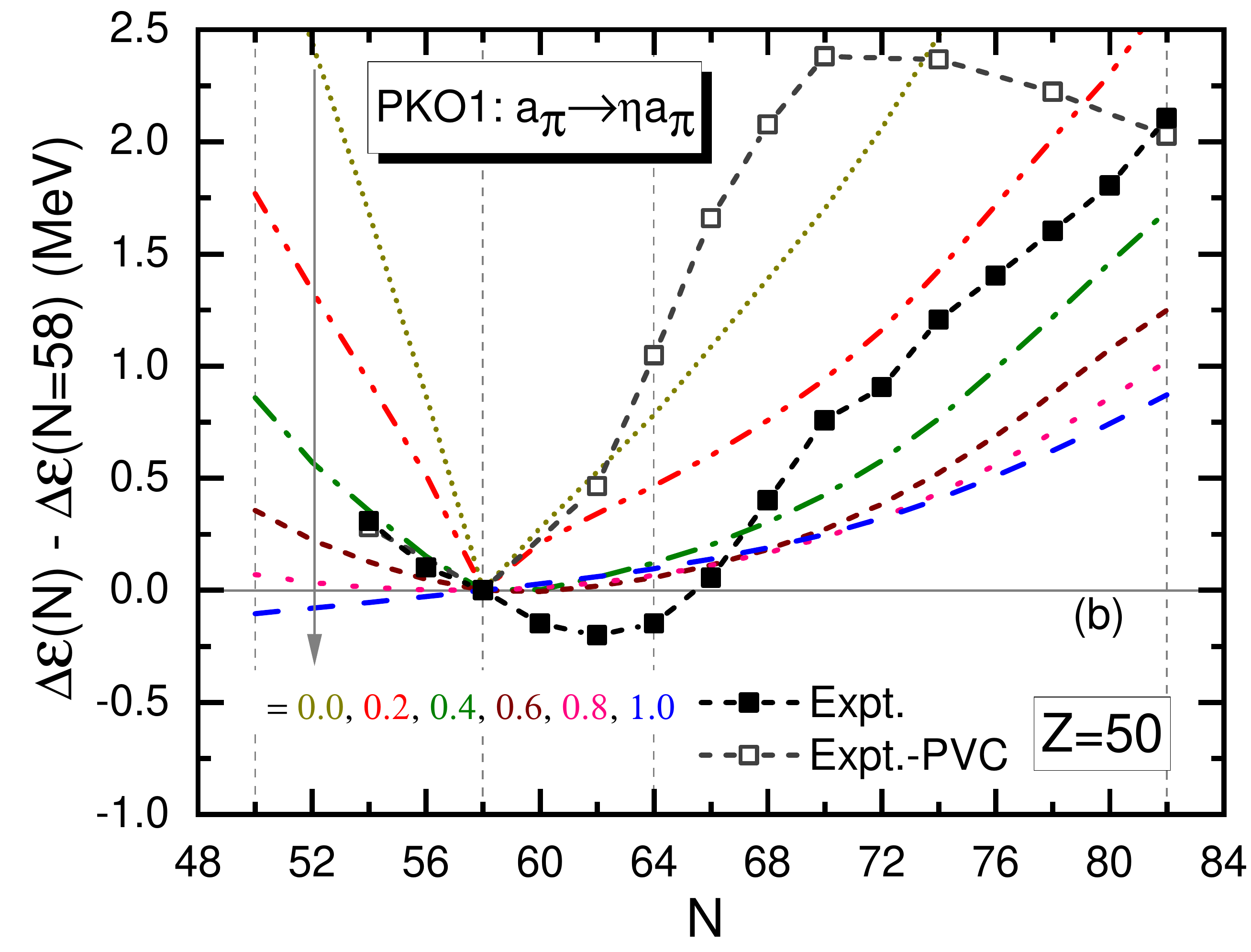}
  \caption{
    Similar to Fig.~\ref{Fig:xfpi}, but with $a_{\pi} \rightarrow \eta a_{\pi}$
    ($ \eta = 1.0 $, $ 0.8 $, $ 0.6 $, $ 0.4 $, $ 0.2 $, and $ 0.0 $).}
  \label{Fig:xapi}
\end{figure*}
\par
Now let us turn to the alternative way to enhance the $\pi$-PV coupling, i.e., reducing the coefficient of density dependence $a_{\pi}$.
In Fig.~\ref{Fig:xapi} are shown the results of PKO1 with
$a_{\pi} \rightarrow \eta a_{\pi}$
($ \eta = 1.0 $, $ 0.8 $, $ 0.6 $, $ 0.4 $, $ 0.2 $, and $ 0.0 $).
One can find that the evolutions of both chains change remarkably with the reduction of $a_{\pi}$ over the whole region discussed here.
When $ \eta \simeq 0.4 $,
the data in the regions of $ N $, $ Z \leq 58 $ are reproduced quite well.
Moreover, for the $N = 82$ isotones ($Z = 50$ isotopes) with
$ 64 \leq \text{$ Z $ ($ N $)} \leq 68 $,
reducing $a_{\pi}$ can also improve the description, whereas increasing $f_{\pi}$ as a whole does not work well.
\par
Based on the discussion above, it is found that enlarging the strength of $\pi$-PV coupling properly can significantly improve the description of the evolution of the single-particle energy difference.
Compared with increasing $f_{\pi}$ with a factor, reducing its density dependence is a preferable way.
It should be noted that we did not perform self-consistent fitting up to this point.
Thus, the optimum values of $\lambda$ and $\eta$ given above should not be taken too seriously, but understood qualitatively.
\par
The weak density dependence of $f_{\pi}$ is related not only to the strength of the tensor force but also to the in-medium effects on it.
It has been argued that the bare tensor force does not undergo significant modification during the renormalization procedure~\cite{
  Otsuka2010Phys.Rev.Lett.104.012501,
  Tsunoda2011Phys.Rev.C84.044322}.
In other words, the effective tensor force in the nuclear medium would be similar with the bare one.
Such a property is known as ``renormalization persistency.''
In the RHF theory with the density-dependent coupling strengths, the renormalization persistency can naturally manifest as weak density dependence. The preference for the small $a_{\pi}$, which is shown above, provides a support for the ``tensor renormalization persistency''~\cite{
  Otsuka2010Phys.Rev.Lett.104.012501,
  Tsunoda2011Phys.Rev.C84.044322}
within the scheme of DFT.
\section{Attempt at a new effective interaction}
\begin{table*}[tb]
  \begin{ruledtabular}
    \caption{
      Parameters of the new effective (denoted as ``New'') interaction within the framework of RHF theory,
      in comparison with those of PKO1~\cite{
        Long2006Phys.Lett.B640.150}.
      The parameters $a_{\phi}$, $b_{\phi}$, $c_{\phi}$, and $d_{\phi}$ are the density dependence coefficients;
      see Ref.~\cite{
        Long2006Phys.Lett.B640.150}
      for the details.}
    \label{tab:new_parameter}
    \begin{tabular}{cl|dddddd}
      & & \multicolumn{1}{r}{$ m_{\phi}$ ($ \mathrm{MeV} $)} & \multicolumn{1}{r}{$g_{\phi} $ or $ f_{\pi} $} & \multicolumn{1}{r}{$a_{\phi} $} & \multicolumn{1}{r}{$b_{\phi}$} & \multicolumn{1}{r}{$c_{\phi} $} & \multicolumn{1}{r}{$d_{\phi} $} \\
      \hline
      \multirow{2}{*}{$ \sigma $}
      & New  & 525.769084 &  9.550360 & 1.294390 & 0.133015 & 0.233251 & 1.195439 \\
      & PKO1 & 525.769084 &  8.833239 & 1.384494 & 1.513190 & 2.296615 & 0.380974 \\
      \hline
      \multirow{2}{*}{$ \omega $}
      & New  & 783.000000 & 11.815491 & 1.375881 & 0.075553 & 0.168960 & 1.404583 \\
      & PKO1 & 783.000000 & 10.729933 & 1.403347 & 2.008719 & 3.046686 & 0.330770 \\
      \hline
      \multirow{2}{*}{$ \rho $}
      & New  & 769.000000 &  2.529138 & 0.684200 &          &          &          \\
      & PKO1 & 769.000000 &  2.629000 & 0.076760 &          &          &          \\
      \hline
      \multirow{2}{*}{$ \pi $}
      & New  & 138.000000 &  1.254333 & 0.659594 &          &          &          \\
      & PKO1 & 138.000000 &  1.000000 & 1.231976 &          &          &          \\
    \end{tabular}
  \end{ruledtabular}
\end{table*}
\begin{table*}[tb]
  \caption{
    Binding energies $E_{\text{b}}$ and charge radii $R_{\text{ch}}$ for the reference nuclei calculated by the new effective interaction (denoted as ``New''),
    in comparison with those calculated by PKO1~\cite{
      Long2006Phys.Lett.B640.150}
    and the corresponding experimental data~\cite{
      Wang2021Chin.Phys.C45.030003,
      Angeli2013At.DataNucl.DataTables99.69}.
    The root-mean-square deviations
    $ \Delta = \sqrt{\sum_{i=1}^N \left( y_i^{\text{Expt.}} - y_i^{\text{Cal.}} \right)^2/N} $
    are also shown.
    }
  \label{Tab:BE_Rch}
  \begin{ruledtabular}
    \begin{tabular}{ldddddd}
      \multirow{2}{*}{Nucleus} & \multicolumn{3}{c}{$E_{\text{b}}$ ($ \mathrm{MeV} $)} & \multicolumn{3}{c}{$R_{\text{ch}}$ ($ \mathrm{fm} $)} \\ \cline{2-7}
                              & \multicolumn{1}{c}{Expt.} & \multicolumn{1}{c}{New} & \multicolumn{1}{c}{PKO1} & \multicolumn{1}{c}{Expt.} & \multicolumn{1}{c}{New} & \multicolumn{1}{c}{PKO1} \\ \hline
      $ \nuc{O}{16}{} $   &   -127.62 &   -127.60   &  -128.15 &  2.699 & 2.668  &  2.676  \\
      $ \nuc{O}{24}{} $   &   -168.96 &   -171.97   &  -170.84 &        &        &         \\
      $ \nuc{S}{36}{} $   &   -308.71 &   -306.00   &  -307.44 &  3.299 & 3.248  &  3.264  \\
      $ \nuc{Ca}{40}{} $  &   -342.05 &   -340.94   &  -342.94 &  3.478 & 3.438  &  3.443  \\
      $ \nuc{Ca}{48}{} $  &   -416.00 &   -418.64   &  -416.98 &  3.477 & 3.439  &  3.451  \\
      $ \nuc{Ca}{52}{} $  &   -438.33 &   -437.28   &  -436.81 &        &        &         \\
      $ \nuc{Ca}{54}{} $  &   -445.37 &   -445.06   &  -445.14 &        &        &         \\
      $ \nuc{Ni}{56}{} $  &   -484.00 &   -480.45   &  -482.63 &        &        &         \\
      $ \nuc{Ni}{68}{} $  &   -590.41 &   -589.88   &  -590.90 &        &        &         \\
      $ \nuc{Ni}{72}{} $  &   -613.46 &   -612.92   &  -614.22 &        &        &         \\
      $ \nuc{Kr}{86}{} $  &   -749.23 &   -751.92   &  -750.81 &  4.184 & 4.157  &  4.166  \\
      $ \nuc{Zr}{90}{} $  &   -783.90 &   -785.00   &  -785.48 &  4.269 & 4.250  &  4.259  \\
      $ \nuc{Ru}{94}{} $  &   -806.86 &   -807.01   &  -809.03 &        &        &         \\
      $ \nuc{Sn}{100}{} $ &   -825.30 &   -825.00   &  -827.83 &        &        &         \\
      $ \nuc{Sn}{116}{} $ &   -988.68 &   -985.99   &  -988.78 &  4.625 & 4.593  &  4.595  \\
      $ \nuc{Sn}{124}{} $ &  -1049.96 &  -1048.06   & -1050.54 &  4.674 & 4.648  &  4.650  \\
      $ \nuc{Sn}{132}{} $ &  -1102.84 &  -1102.96   & -1103.00 &  4.709 & 4.700  &  4.705  \\
      $ \nuc{Xe}{136}{} $ &  -1141.88 &  -1145.98   & -1145.57 &  4.796 & 4.793  &  4.796  \\
      $ \nuc{Ce}{140}{} $ &  -1172.68 &  -1180.18   & -1177.97 &  4.877 & 4.870  &  4.875  \\
      $ \nuc{Gd}{146}{} $ &  -1204.43 &  -1206.82   & -1208.68 &  4.980 & 4.984  &  4.985  \\
      $ \nuc{Pb}{182}{} $ &  -1411.65 &  -1412.57   & -1416.47 &  5.379 & 5.378  &  5.376  \\
      $ \nuc{Pb}{194}{} $ &  -1525.89 &  -1523.40   & -1527.14 &  5.437 & 5.445  &  5.449  \\
      $ \nuc{Pb}{200}{} $ &  -1576.36 &  -1573.97   & -1576.65 &  5.461 & 5.484  &  5.483  \\
      $ \nuc{Pb}{204}{} $ &  -1607.51 &  -1605.84   & -1607.66 &  5.480 & 5.503  &  5.504  \\
      $ \nuc{Pb}{208}{} $ &  -1636.43 &  -1635.85   & -1636.76 &  5.501 & 5.521  &  5.521  \\
      $ \nuc{Pb}{214}{} $ &  -1663.29 &  -1662.54   & -1663.45 &  5.558 & 5.587  &  5.584  \\
      $ \nuc{Po}{210}{} $ &  -1645.21 &  -1647.65   & -1648.54 &  5.570 & 5.553  &  5.551  \\
      $ \nuc{Ra}{214}{} $ &  -1658.32 &  -1665.25   & -1665.51 &  5.608 & 5.619  &  5.614  \\
      $ \nuc{U}{218}{} $  &  -1665.68 &  -1675.94   & -1674.94 &        &        &         \\
      \hline
      $\Delta$            &           &      3.28   &     3.00 &        & 0.024  & 0.020   \\
    \end{tabular}
  \end{ruledtabular}
\end{table*}
\begin{figure*}
  \includegraphics[width=0.45\textwidth]{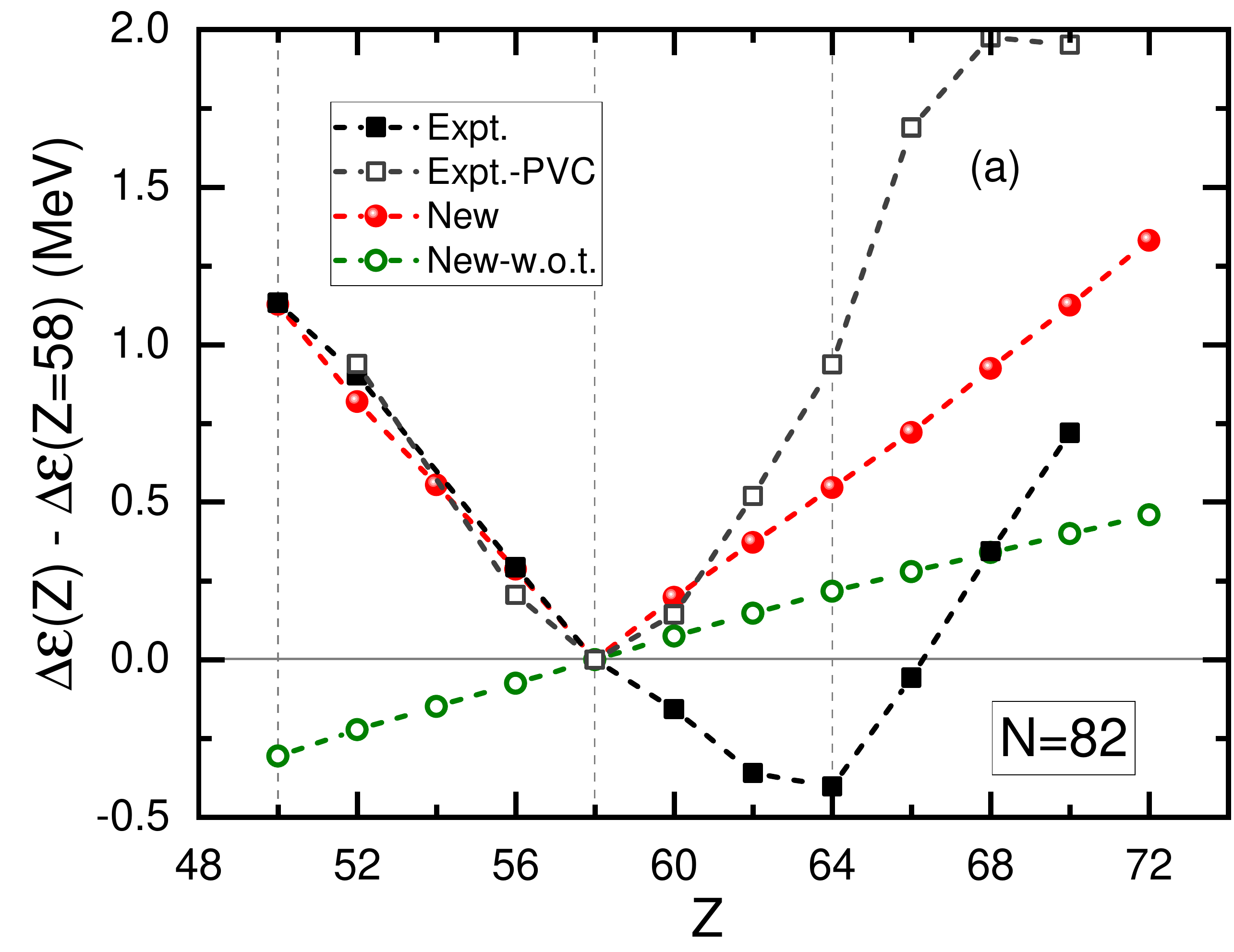}
  \hspace{0.5cm}
  \includegraphics[width=0.45\textwidth]{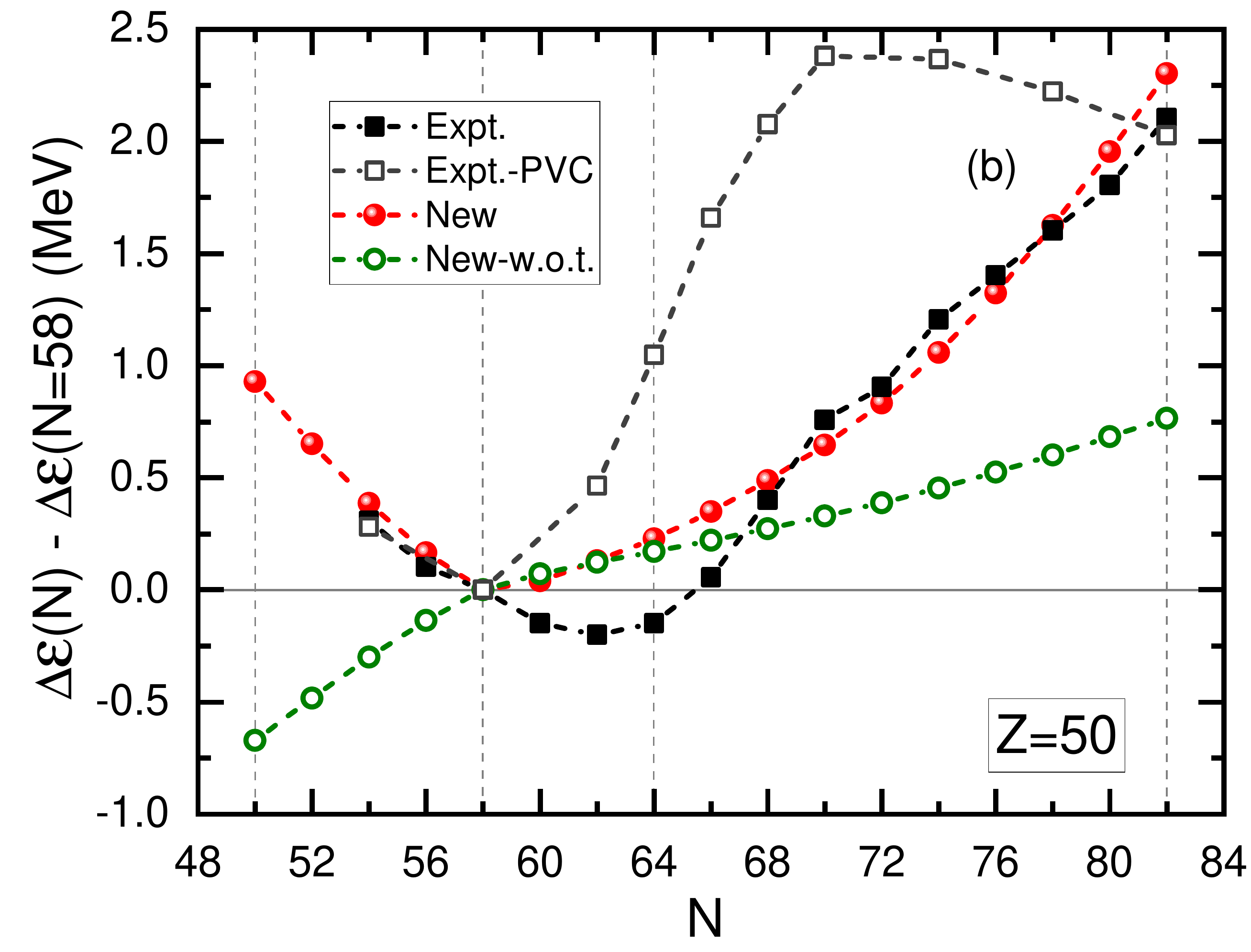}
  \caption{
    (a) Energy differences
    $\Delta \varepsilon \equiv \varepsilon_{\nu 1i_{13/2}}-\varepsilon_{\nu 1h_{9/2}}$
    in the $N=82$ isotones as functions of the proton number,
    normalized with respect to the values at $Z = 58$.
    (b) Energy differences
    $\Delta \varepsilon \equiv \varepsilon_{\pi 1h_{11/2}}-\varepsilon_{\pi 1g_{7/2}}$
    in the $Z=50$ isotopes as functions of the neutron number,
    normalized with respect to the values at $N = 58$.
    The calculation is performed by the RHF theory with the new effective interaction.
    For comparison, the experimental data are also given, which are the same as those in Figs.~\ref{Fig:N82-pairing}(a) and \ref{Fig:Z50-pairing}(a).}
  \label{Fig:Z=82-N=50-New}
\end{figure*}
\par
In the previous section, it is found that the description of the shell-structure evolution can be improved by enhancing the strength of the tensor force.
However, one may notice that the readjustment of the strength of the $\pi$-PV coupling (essentially the strength of the tensor force) is not performed in a self-consistent manner, which may call into question the reliability of the conclusion above.
Thus, it is meaningful to develop a new set of effective interaction within the framework of RHF theory under the guidance of the foregoing discussion.
\par
We start from PKO1, aiming at a new effective interaction with the same meson-nucleon couplings and better-constrained tensor force.
In the fitting procedure, we find that one is unlikely to achieve a satisfying functional by enlarging $f_{\pi} \left( 0 \right)$ or weakening $a_{\pi}$ individually.
In practice, the initial value of $a_{\pi}$ is set very small in order to get a relatively small final value.
The parameters are, as an attempt, fitted to the bulk properties of the nuclear matter and several selected nuclei.
A new effective interaction with larger $f_{\pi} \left( 0 \right)$ ($\lambda \simeq 1.25$) and smaller $a_{\pi}$ ($\eta \simeq 0.54$), in contrast to PKO1, is obtained.
The parameters of the new effective interaction are displayed in Table~\ref{tab:new_parameter}.
For the new effective  interaction, the strength of the $\pi$-PV coupling at the saturation density $f_{\pi} \left(\rho_{\text{sat.}}\right)$ is $ 0.649 $, while it is $ 0.293 $ for PKO1.
Accordingly, one can find that the tensor force in the new effective interaction is significantly (roughly twice) stronger than that in PKO1.
\par
We have calculated the binding energies and charge radii of a group of nuclei ranging from light to heavy ones. The results are shown in Table~\ref{Tab:BE_Rch}, in comparison with those calculated by PKO1. As can be seen, the present effective interaction can describe the bulk properties of the finite nuclei almost as accurately as PKO1 does.
\par
In addition, the energy differences between the single-neutron states $\nu 1i_{13/2}$ and $\nu 1h_{9/2}$ along the $N=82$ isotones and those between the single-proton states $\pi 1h_{11/2}$ and $\pi 1g_{7/2}$ along the $Z = 50$ isotopes are calculated with the new effective interaction.
The results of the $N=82$ isotones and the $Z = 50$ isotopes are shown in Figs.~\ref{Fig:Z=82-N=50-New}(a) and~\ref{Fig:Z=82-N=50-New}(b), respectively, in comparison with the corresponding experimental data.
One can see that the energy differences in each chain, which are denoted by the red filled circles, present a remarkable turning point.
More importantly, the slopes on both sides of the turning point are enhanced dramatically and get much closer to the data, compared with the results of previous RHF functionals.
Thus, it is clearly shown that the description of the shell-structure evolution is significantly improved by the new effective interaction.
\par
Since the most distinct difference of the new effective interaction from the previous PKO$i$ ($ i = 1 $, $ 2 $, $ 3 $) is its enhanced strength of $\pi$-PV coupling, it is reasonable to suspect that the improvement of the description of the shell-structure evolution is mainly due to the relatively stronger tensor force in it.
To prove this, we calculate the contribution of the tensor force to the energy differences and subtract it from the results of the full calculation.
The corresponding results, denoted by the green open circles, are also shown in Figs.~\ref{Fig:Z=82-N=50-New}(a) and~\ref{Fig:Z=82-N=50-New}(b), respectively.
Without the contribution of the tensor force, the energy differences calculated by the new effective interaction increase monotonically with the proton and neutron numbers.
Clearly, one can conclude that the ability of the new effective interaction to improve the description of the shell-structure evolution is attributed to its strong tensor force.
\par
Through self-consistent refitting, it is shown that enhancing the strength of the tensor force in the RHF effective interaction can improve the description of the shell-structure evolution, without considerable sacrifice of the accuracy on the bulk properties.
Meanwhile, reducing the coefficient of the density dependence $a_{\pi}$, which is related to the idea of tensor renormalization persistency~\cite{
  Otsuka2010Phys.Rev.Lett.104.012501,
  Tsunoda2011Phys.Rev.C84.044322},
is shown to be practicable.
\section{Summary and Perspectives}
\label{sec:summary}
\par
Within the framework of the RHF theory, the evolutions of the energy difference between the single-neutron states $\nu 1i_{13/2}$ and $\nu 1h_{9/2}$ along the $N=82$ isotonic chain and that between the single-proton states $\pi 1h_{11/2}$ and $\pi 1g_{7/2}$ along the $Z = 50$ isotopic chain have been investigated.
Our main focus is on the effects of the tensor force.
To compare the calculated results with the experimental data as informative as possible, we have adopted not only the original data but also the pseudodata in which the PVC correlation is removed.
It is found that the tensor force plays crucial roles in properly describing the shell-structure evolutions.
The contributions of the tensor force from each relevant meson-nucleon coupling were studied in details.
It is shown that, for PKA1, PKO1, and PKO3, the tensor force from the $\pi$-PV coupling plays a decisive role, while for PKO2, which does not explicitly contain the $\pi$-PV coupling, the net contribution of the tensor force is negligible.
By comparing with the data, we find that the strength of the tensor force is of vital importance for (qualitatively) reproducing the trend of the data.
\par
In addition, the strength of the tensor force in the CDFT has been further explored.
It is found that moderately increasing the coupling strength of $\pi$-PV coupling, which essentially enhances the tensor force in the effective interaction, will significantly improve the description of the shell-structure evolution.
A systematic comparison shows that weakening the density dependence of the $\pi$-PV coupling is a more efficient way than enlarging it with a factor.
Our study thus provides a support for the idea of tensor renormalization persistency~\cite{
  Otsuka2010Phys.Rev.Lett.104.012501,
  Tsunoda2011Phys.Rev.C84.044322},
which emphasizes the resemblance between the in-medium effective tensor force and the bare one.
Moreover, guided by the exploration of the tensor force, we have attempted to develop a new effective interaction with stronger $\pi$-PV coupling, which is mainly determined by the weakened density dependence (smaller $a_{\pi}$) and partially by the enhanced strength at zero density (larger $f_{\pi} \left( 0 \right)$).
The new effective interaction can improve the description of the shell-structure significantly, without considerable sacrifice of the accuracy on the bulk properties.
It has also been shown that the improvement is attributed to the enhancement of the tensor force.
\par
As a perspective, it is meaningful to develop the framework of PVC based on the RHF theory in the future, so as to treat the PVC correlation in a self-consistent way.
An alternative way to avoid the distraction of the dynamical correlation is to seek the metadata which belong to the pure mean-field level.
The spin-orbit splitting calculated by the relativistic Brueckner-Hartree-Fock theory~\cite{
  Shen2016Chin.Phys.Lett.33.102103,
  Shen2017Phys.Rev.C96.014316,
  Shen2019Prog.Part.Nucl.Phys.109.103713}
has attracted great attention~\cite{
  Shen2018Phys.Lett.B778.344,
  Shen2018Phys.Lett.B781.227,
  Shen2019Phys.Rev.C99.034322,
  Wang2019Phys.Rev.C100.064319,
  Zhao2020Phys.Rev.C102.034322,
  Ge2020Phys.Rev.C102.044304}.
Work in this direction is also in progress.
\begin{acknowledgments}
  This work is partly supported by the Natural Science Foundation of China under Grants No.~11905088, No.~11675065, and No.~12075104, by the MOST ``Introduction Plan Program of High-Level Foreign Talents'' under Grant No.~G20200028008, by the JSPS Grant-in-Aid for Early-Career Scientists under Grant No.~18K13549, by the JSPS Grant-in-Aid for JSPS Fellows under Grant No.~19J20543, and by the JSPS Grant-in-Aid for Scientific Research (S) under Grant No.~20H05648.
  Z.W. acknowledges support by a scholarship from the China Scholarship Council.
  T.N. and H.L. thank the RIKEN iTHEMS program and the RIKEN Pioneering Project:~Evolution of Matter in the Universe.
\end{acknowledgments}
\end{document}